\documentstyle[preprint,prd,aps]{revtex}
\def\Cadez{\v{C}ade\v{z}}
\def\Masso{Mass\'{o}}
\def\OMurchadha{\`{O}~Murchadha}
\def\Strahlkorper{Strahl\-k\"{o}rper}		% Minkowski's term
\def\csmash#1{\hbox to 0em{\hss{#1}\hss}}	% center and smash
\def\defn#1{``#1''}
\def\del{\nabla}
\def\gtsim{\gtrsim}

\def\interp{\mathop{\rm interp}\nolimits}	% defined the same way as all
\def\ltsim{\lesssim}
\def\tfrac#1#2{{\textstyle \frac{#1}{#2}}}
\def\thalf{\tfrac{1}{2}}
\def\A{{\cal A}}		% 2-D (continuous) domain of angular coords
\def\N{{\cal N}}		% 3-D (continuous) neighborhood of horizon
\def\Jacc[#1]{J \Bigl[ #1 \Bigr]}		% continuum Jacobian for fn #1
\def\Jac[#1]{{\bf J} \Bigl[ #1 \Bigr]}		% discrete Jacobian for fn #1
\def\e{{\sf e}}					% grid function $e$
\def\h{{\sf h}}					% grid function $h$
\def\H{{\sf H}}					% grid function $H$
\def\n{{\sf n}}					% grid function $n$
\def\P{{\sf P}}					% grid function $P$
\def\Q{{\sf Q}}					% grid function $Q$
\def\s{{\sf s}}					% grid function $s$
\def\I{{\text{\scriptsize I}}}			% grid-point index
\def\J{{\text{\scriptsize J}}}			% grid-point index
\def\K{{\text{\scriptsize K}}}			% grid-point index
\def\M{{\text{\scriptsize M}}}			% molecule index
\def\dd{\text{\bf d}}				% generic derivative molecule
\def\II{\text{\bf I}}				% identity finite diff molecule
\def\MM{\text{\sf M}}				% generic finite diff molecule
\def\two{{}^{(2)}}				% (2) prefix superscript
\def\three{{}^{(3)}}				% (3) prefix superscript
\def\booktitle#1{{\it #1\/}}
\def\papertitle#1{{\it #1\/}}
\def\reporttitle#1{{\it #1\/}}
\def\talktitle#1{{\it #1\/}}
\def\thesistitle#1{{\it #1\/}}
\def\eqref#1{Eq.~$(\text{\ref{#1}})$}
\def\eqrefs#1{Eqs.~$(\text{\ref{#1}})$}
\def\eqrefb#1{$(\text{\ref{#1}})$}		% "b" = "bare"
\tighten			% single spacing
\begin{document}
\draft				% print PACS numbers
\preprint{gr-qc/9508014}
\title{Finding Apparent Horizons in Numerical Relativity}
\author{Jonathan Thornburg%%%
	\thanks{Address for written correspondence:
		Box~8--7, Thetis Island,
		British Columbia, V0R~2Y0, Canada}}
\address{Physics Department, University of British Columbia	\\
	 Vancouver, British Columbia, V6T~1Z1, Canada		\\
	 \vbox{\tt thornbur@theory.physics.ubc.ca}}
\date{\today}
\maketitle
%
%%%%%%%%%%%%%%%%%%%%%%%%%%%%%%%%%%%%%%%%%%%%%%%%%%%%%%%%%%%%%%%%%%%%%%%%%%%%%%%
%
\begin{abstract}
We review various algorithms for finding apparent horizons in $3+1$
numerical relativity.  We then focus on one particular algorithm,
in which we pose the apparent horizon equation
$H \equiv \del_i n^i + K_{ij} n^i n^j - K = 0$
as a nonlinear elliptic (boundary-value) PDE on angular-coordinate
space for the horizon shape function $r = h(\theta,\phi)$, finite
difference this PDE, and use Newton's method or a variant to solve
the finite difference equations.

We describe a method for computing the Jacobian matrix of the finite
differenced $H(h)$ function $\H(\h)$ by symbolically differentiating
the finite difference equations, giving the Jacobian elements directly
in terms of the finite difference molecule coefficients used in computing
$\H(\h)$.  Assuming the finite differencing scheme commutes with
linearization, we show how the Jacobian elements may be computed by
first linearizing the continuum $H(h)$ equations, then finite differencing
the linearized continuum equations.  (This is essentially just the
``Jacobian part'' of the Newton-Kantorovich method for solving nonlinear
PDEs.)  We tabulate the resulting Jacobian coefficients for a number
of different $\H(\h)$ and Jacobian computation schemes.

We find this symbolic differentiation method of computing the
$\H(\h)$ Jacobian to be {\em much\/} more efficient than the usual
numerical-perturbation method, and also much easier to implement
than is commonly thought.

When solving the discrete $\H(\h) = 0$ equations, we find that
Newton's method generally shows robust convergence.  However, we find
that it has a small (poor) radius of convergence if the initial guess
for the horizon position contains significant high-spatial-frequency
error components, i.e.~angular Fourier components varying as (say)
$\cos m\theta$ with $m \gtsim 8$.  (Such components occur naturally
if spacetime contains significant amounts of high-frequency gravitational
radiation.)  We show that this poor convergence behavior is {\em not\/}
an artifact of insufficient resolution in the finite difference grid;
rather, it appears to be caused by a strong nonlinearity in the continuum
$H(h)$ function for high-spatial-frequency error components in $h$.
We find that a simple ``line search'' modification of Newton's method
roughly doubles the horizon finder's radius of convergence, but both
the unmodified and modified methods' radia of convergence still fall
rapidly with increasing spatial frequency, approximately as $1 / m^{3/2}$.
Further research is needed to explore more robust numerical algorithms
for solving the $\H(\h) = 0$ equations.

Provided it converges, the Newton's-method algorithm for horizon
finding is potentially very accurate, in practice limited only by
the accuracy of the $\H(\h)$ finite differencing scheme.  Using
4th~order finite differencing, we demonstrate that the error in
the numerically computed horizon position shows the expected
$O((\Delta \theta)^4)$ scaling with grid resolution $\Delta \theta$,
and is typically $\sim \! 10^{-5} (10^{-6})$ for a grid resolution of
$\Delta \theta = \frac{\pi/2}{50} (\frac{\pi/2}{100})$.

Finally, we briefly discuss the global problem of finding or recognizing
the {\em outermost\/} apparent horizon in a slice.  We argue that
this is an important problem, and that no reliable algorithms currently
exist for it except in spherical symmetry.
\end{abstract}
%
%%%%%%%%%%%%%%%%%%%%%%%%%%%%%%%%%%%%%%%%%%%%%%%%%%%%%%%%%%%%%%%%%%%%%%%%%%%%%%%
%
%
%%%%%%%%%%%%%%%%%%%%%%%%%%%%%%%%%%%%%%%%%%%%%%%%%%%%%%%%%%%%%%%%%%%%%%%%%%%%%%%
%
\section{Introduction}
\label{sect-introduction}

In $3+1$ numerical relativity, one often wishes to locate the black
hole(s) in a (spacelike) slice.  As discussed by Refs.~\cite{Hawking-73,%%%
Hawking-Ellis}, a black hole is rigorously defined in terms of its
event horizon, the boundary of future null infinity's causal past.
Although the event horizon has, in the words of Hawking and Ellis
(Ref.~\cite{Hawking-Ellis-quote}), ``a number of nice properties'',
it's defined in an inherently {\em acausal\/} manner: it can only be
determined if the entire future development of the slice is known.
(As discussed by Refs.~\cite{ABBLMSSSW-94,LMSSW-95}, in practice the
event horizon may be located to good accuracy given only the usual
numerically generated approximate development to a nearly stationary
state, but the fundamental acausality remains.)

In contrast, an apparent horizon, also known as a marginally outer
trapped surface, is defined (Refs.~\cite{Hawking-73,Hawking-Ellis})
locally in time, within a single slice, as a closed 2-surface whose
outgoing null geodesics have zero expansion.  An apparent horizon
is slicing-dependent: if we define a ``world tube'' by taking the
union of the apparent horizon(s) in each slice of a slicing, this
world tube will vary from one slicing to another.  In a stationary
spacetime event and apparent horizons coincide, although this generally
isn't the case in dynamic spacetimes.  However, given certain technical
assumptions, the existence of an apparent horizon in a slice implies
the existence of an event horizon, and thus by definition a black
hole, containing the apparent horizon.  (Unfortunately, the converse
doesn't always hold.  Notably, Wald and Iyer (Ref.~\cite{Wald-Iyer-91})
have constructed a family of angularly anisotropic slices in Schwarzschild
spacetime which approach arbitrarily close to $r = 0$ yet contain no
apparent horizons.)

There is thus considerable interest in numerical algorithms to find
apparent horizons in numerically computed slices, both as diagnostic
tools for locating black holes and studying their behavior (see, for
example, Refs.~\cite{ABBLMSSSW-94,ABBHSS-94}), and for use ``on the fly''
during numerical evolutions to help in choosing the coordinates and
``steering'' the numerical evolution (Refs.~\cite{Thornburg-talk-91,%%%
Seidel-Suen-92,Thornburg-PhD,ADMSS-95}).  This latter context makes
particularly strong demands on a horizon-finding algorithm:  Because
the computed horizon position is used in the coordinate conditions,
the horizon must be located quite accurately to ensure that spurious
finite difference instabilities don't develop in the time evolution.
Furthermore, the horizon must be re-located at each time step of the
evolution, so the horizon-finding algorithm should be as efficient as
possible.  Finally, when evolving multiple-black-hole spacetimes in
this manner it's desirable to have a means of detecting the appearance
of a new outermost apparent horizon surrounding two black holes which
are about to merge.  We discuss this last problem further in
section~\ref{sect-finding-outermost-apparent-horizons}.

In this paper we give a detailed discussion of the \defn{Newton's method}
apparent-horizon-finding algorithm.  This algorithm poses the
apparent horizon equation as a nonlinear elliptic (boundary-value)
PDE on angular-coordinate space for the horizon shape function
$r = h(\theta,\phi)$, finite differences this PDE, and uses some
variant of Newton's method to solve the resulting set of simultaneous
nonlinear algebraic equations for the values of $h$ at the
angular-coordinate grid points.  This algorithm is suitable for both
axisymmetric and fully-general spacetimes, and we discuss both cases.
As explained in section~\ref{sect-notation}, we assume a locally
polar spherical topology for the coordinates and finite differencing,
though we make no assumptions about the basis used in taking tensor
components.
%
%%%%%%%%%%%%%%%%%%%%%%%%%%%%%%%%%%%%%%%%%%%%%%%%%%%%%%%%%%%%%%%%%%%%%%%%%%%%%%%
%
\section{Notation}
\label{sect-notation}

Our notation generally follows that of Misner, Thorne, and Wheeler
(Ref.~\cite{MTW}), with $G = c = 1$ units and a $(-,+,+,+)$ spacetime
metric signature.  We assume the usual Einstein summation convention
for repeated indices regardless of their tensor character, and we use
the Penrose abstract-index notation, as described by (for example)
Ref.~\cite{Wald}.  We use the standard $3+1$ formalism of Arnowitt,
Deser, and Misner (Ref.~\cite{ADM-62}) (see Refs.~\cite{York-79,%%%
York-83} for recent reviews).

We assume that a specific spacetime and $3+1$ (spacelike) slice are
given, and all our discussions take place within this slice.  We use
the term \defn{horizon} to refer to the (an) apparent horizon in this
slice.  We often refer to various sets in the slice as being 1, 2,
or 3-dimensional, meaning the number of {\em spatial\/} dimensions --
the time coordinate is never included in the dimensionality count.
For example, we refer to the horizon itself as 2-dimensional.

We assume that the spatial coordinates $x^i \equiv (r,\theta,\phi)$
are such that in some neighborhood of the horizon, surfaces of constant
$r$ are topologically nested 2-spheres with $r$ increasing outward,
and we refer to $r$ as a \defn{radial} coordinate and $\theta$ and
$\phi$ as \defn{angular} coordinates.  For pedagogical convenience
(only), we take $\theta$ and $\phi$ to be the usual polar spherical
coordinates, so that if spacetime is axisymmetric (spherically symmetric),
$\phi$ is ($\theta$ and $\phi$ are) the symmetry coordinate(s).  However,
we make no assumptions about the detailed form of the coordinates,
i.e.~we allow all components of the 3-metric to be nonzero.

We emphasize that although our assumptions about the local topology
of $r$ are fundamental, our assumptions about the angular coordinates
are for pedagogical convenience only, and could easily be eliminated.
In particular, all our discussions carry over unchanged to multiple
black hole spacetimes, using (for example) either \Cadez{}
conformal-mapping equipotential coordinates (Ref.~\cite{Cadez-PhD})
or multiple-coordinate-patch coordinate systems (Ref.~\cite{Thornburg-87}).

We use $ijkl$ for spatial (3-) indices, and $uvwxy$ for indices
ranging over the angular coordinates only.  $g_{ij}$ denotes the
3-metric in the slice, $g$ its determinant, and $\del_i$ the associated
3-covariant derivative operator.  $K_{ij}$ denotes the 3-extrinsic
curvature of the slice, and $K$ its trace.

We use $\A$ to denote the 2-dimensional space of angular coordinates
$(\theta,\phi)$.  We sometimes need to distinguish between field
variables defined on $\A$ or on the (2-dimensional) horizon, and
field variables defined on a 3-dimensional neighborhood $\N$ of
the horizon.  This distinction is often clear from context, but
where ambiguity might arise we use prefixes $\two$ and $\three$
respectively, as in $\two\! H$ and $\three\! H$.

We use italic Roman letters $H$, $h$, etc., to denote {\em continuum\/}
coordinates, functions, differential operators, and other quantities.
We use sans serif Roman letters $\H$, $\h$, etc., to denote grid
functions, and small capital Roman indices $\I$, $\J$, and $\K$ to
index grid points.  We use subscript grid-point indices to denote
the evaluation of a continuum or grid function at a particular grid
point, as in $H_\I$ or $\H_\I$.  We use $\Jac[\P(\Q)]$ to denote the
Jacobian matrix of the grid function $\P = \P(\Q)$, as defined by
\eqref{eqn-Jac[P(Q)]}, and ${} \cdot {}$ to denote the product of
two such Jacobians or that of a Jacobian and a grid function.  We
use $\Jacc[P(Q)]$ to denote the linearization of the differential
operator $P = P(Q)$ about the point $Q$.

We use $\MM$ as a generic finite difference molecule and $\M$ as
a generic index for molecule coefficients.  We write $\M \in \MM$
to mean that $\MM$ has a nonzero coefficient at position $\M$.
Temporarily taking $\langle \M \rangle$ to denote some particular
coordinate component of $\M$, we refer to
$\max_{\M \in \MM} | \langle \M \rangle |$ as the \defn{radius} of
$\MM$, and to the number of distinct $\langle \M \rangle$ values
with $\M \in \MM$ as the \defn{diameter} or \defn{number of points}
of $\MM$.  (For example, the usual symmetric 2nd~order 3-point molecules
for 1st~and 2nd~derivatives both have radius~1 and diameter~3.)  We
often refer to a molecule as itself being a discrete operator, the
actual application to a grid function being implicit.

%
% FIXME:
%	It would be nicer to use \; instead of \,\, for the
%	spacing around $\interp(...)$ in this paragraph.  Alas,
%	LaTeX doesn't allow \; outside of math mode (sigh)...
%
Given a grid function $f$ and a set of points $\{ x_k \}$ in its
domain, we use \,\,$\interp(f(x), x=a)$\,\, to mean an interpolation
of the values $f(x_k)$ to the point $x=a$ and \,\,$\interp'(f(x), x=a)$\,\,
to mean the derivative of the same interpolant at this point.  More
precisely, taking $I$ to be a smooth interpolating function (typically
a Lagrange polynomial) such that $I(x_k) = f(x_k)$ for each $k$,
\,\,$\interp(f(x), x=a)$\,\, denotes $I(a)$ and \,\,$\interp'(f(x), x=a)$\,\,
denotes $\Bigl. (\partial I / \partial x) \Bigr|_{x=a}$.
%
%%%%%%%%%%%%%%%%%%%%%%%%%%%%%%%%%%%%%%%%%%%%%%%%%%%%%%%%%%%%%%%%%%%%%%%%%%%%%%%
%
\section{The Apparent Horizon Equation}
\label{sect-apparent-horizon-equation}

As discussed by (for example) Ref.~\cite{York-89}, an apparent
horizon satisfies the equation
\begin{equation}
H \equiv \two\! H \equiv
	\del_i n^i + K_{ij} n^i n^j - K = 0
					\, \text{,}	% text punctuation
							\label{eqn-horizon}
\end{equation}
where $n^i$ is the outward-pointing unit normal to the horizon,
all the field variables are evaluated on the horizon surface, and
where for future use we define the \defn{horizon function}
$H \equiv \two\! H$ as the left hand side of~\eqref{eqn-horizon}.
(Notice that in order for the 3-divergence $\del_i n^i$ to be
meaningful, $n^i$ must be (smoothly) continued off the horizon, and
extended to a field $\three n^i$ in some 3-dimensional neighborhood
of the horizon.  The off-horizon continuation is non-unique, but
it's easy to see that this doesn't affect $H$ on the horizon.)

To solve the apparent horizon equation~\eqref{eqn-horizon}, we
begin by assuming that the horizon and coordinates are such that
each radial coordinate line $\{ (\theta,\phi)=\text{constant} \}$
intersects the horizon in exactly one point.  In other words, we
assume that the horizon's coordinate shape is a \defn{\Strahlkorper},
defined by Minkowski as ``a region in $n$-dimensional Euclidean
space containing the origin and whose surface, as seen from the
origin, exhibits only one point in any direction''
(Ref.~\cite{Schroeder-quote}).  Given this assumption, we can
parameterize the horizon's shape by $r = h(\theta,\phi)$ for some
single-valued \defn{horizon shape function} $h$ defined on the
2-dimensional domain $\A$ of angular coordinates $(\theta,\phi)$.

Equivalently, we may write the horizon's shape as $\three\! F = 0$,
where the scalar function $\three\! F$, defined on some 3-dimensional
neighborhood $\N$ of the horizon, satisfies $\three\! F = 0$ if and
only if $r = h(\theta,\phi)$, and we take $\three\! F$ to increase
outward.  In practice we take
$\three\! F(r,\theta,\phi) = r - h(\theta,\phi)$.

We define the non-unit outward-pointing normal (field) to the horizon
by
\begin{equation}
s_i \equiv \three\! s_i = \del_i \three\! F
					\, \text{,}	% text punctuation
\end{equation}
i.e.~by
\begin{mathletters}
							\label{eqn-s-d(h)}
\begin{eqnarray}
s_r	& = &	1							\\
s_u	& = &	- \partial_u h
					\, \text{.}	% text punctuation
									%%%\\
\end{eqnarray}
\end{mathletters}
and the outward-pointing unit normal (field) to the horizon by
\begin{eqnarray}
n^i \equiv \three n^i
	& = &	\frac{s^i}{\|s^k\|}					\\
	& = &	\frac{g^{ij} s_j}{\sqrt{g^{kl} s_k s_l}}
							\label{eqn-n-u(s-d)}
									\\
	& = &	\frac{g^{ir} - g^{iu} \partial_u h}
		     { \sqrt{ g^{rr}
			      - 2 g^{ru} \partial_u h
			      + g^{uv} (\partial_u h) (\partial_v h) } }
					\, \text{.}	% text punctuation
							\label{eqn-n-u(h)}
									%%%\\
\end{eqnarray}
Henceforth we drop the $\three$ prefixes on $\three\! s_i$ and
$\three n^i$.

Substituting \eqref{eqn-n-u(h)} into the apparent horizon
equation~\eqref{eqn-horizon}, we see that the horizon function
$H(h)$ depends on the (angular) 2nd derivatives of $h$.  In fact,
the apparent horizon equation~\eqref{eqn-horizon} is a 2nd~order
elliptic (boundary-value) PDE for $h$ on the domain of angular
coordinates $\A$.  The apparent horizon equation~\eqref{eqn-horizon}
must therefore be augmented with suitable boundary conditions to
define a (locally) unique solution.  These are easily obtained by
requiring the horizon's 3-dimensional shape to be smooth across the
artificial boundaries $\theta = 0$, $\theta = \pi$, $\phi = 0$, and
$\phi = 2\pi$.
%
%%%%%%%%%%%%%%%%%%%%%%%%%%%%%%%%%%%%%%%%%%%%%%%%%%%%%%%%%%%%%%%%%%%%%%%%%%%%%%%
%
\section{Algorithms for Solving the Apparent Horizon Equation}
\label{sect-algorithms-survey}

We now survey various algorithms for solving the apparent horizon
equation~\eqref{eqn-horizon}.  Ref.~\cite{Nakamura-Oohara-Kojima-87}
reviews much of the previous work on this topic.

In spherical symmetry, the apparent horizon equation~\eqref{eqn-horizon}
degenerates into a 1-dimensional nonlinear algebraic equation for
the horizon radius $h$.  This is easily solved by zero-finding on
the horizon function $H(h)$.  This technique has been used by a number
of authors, for example Refs.~\cite{Petrich-Shapiro-Teukolsky-85,%%%
Choptuik-PhD,Seidel-Suen-92,ADMSS-95}.  (See also
Ref.~\cite{Bizon-Malec-OMurchadha-88} for an interesting analytical
study giving necessary and sufficient conditions for apparent
horizons to form in non-vacuum spherically symmetric spacetimes.)

In an axisymmetric spacetime, the angular-coordinate space $\A$ is
effectively 1-di\-men\-sional, so the apparent horizon
equation~\eqref{eqn-horizon} reduces to a nonlinear 2-point boundary
value (ODE) problem for the function $h(\theta)$, which may be solved
either with a shooting method, or with one of the more general methods
described below.  Shooting methods have been used by a number of
authors, for example Refs.~\cite{Cadez-74,Dykema-PhD,%%%
Abrahams-Evans-92,Bishop-82,Bishop-84,Shapiro-Teukolsky-92,%%%
Abrahams-Heiderich-Shapiro-Teukolsky-92}.

The remaining apparent-horizon-finding algorithms we discuss are all
applicable to either axisymmetric spacetimes (2-dimensional codes) or
fully general spacetimes (3-dimensional codes).

Tod (Ref.~\cite{Tod-91}) has proposed an interesting pair of
\defn{curvature flow} methods for finding apparent horizons.
Bernstein (Ref.~\cite{Bernstein-93}) has tested these methods in
several axisymmetric spacetimes, and reports favorable results.
Unfortunately, the theoretical justification for these methods'
convergence is only valid in time-symmetric ($K_{ij} = 0$) slices.

The next two algorithms we discuss are both based on a pseudospectral
expansion of the horizon shape function $h(\theta,\phi)$ in some
complete set of basis functions (typically spherical harmonics or
symmetric trace-free tensors), using some finite number of the expansion
coefficients $\{ a_k \}$ to parameterize of the horizon shape.  One
algorithm rewrites the apparent horizon equation $H(a_k) = 0$ as
$\|H(a_k)\| = 0$, then uses a general-purpose function-minimization
routine to search $\{ a_k \}$-space for a minimum of $\|H\|$.  This
algorithm has been used by Refs.~\cite{Brill-Lindquist-63,Eppley-77}
in axisymmetric spacetimes, and more recently by
Ref.~\cite{Libson-Masso-Seidel-Suen-95} in fully general spacetimes.
Alternatively, Nakamura, Oohara, and Kojima
(Refs.~\cite{Nakamura-Kojima-Oohara-84,Oohara-Nakamura-Kojima-85,%%%
Oohara-86}) have suggested a functional iteration method for directly
solving the apparent horizon equation $H(a_k) = 0$ for the expansion
coefficients $\{ a_k \}$, and have used it in a number of fully general
spacetimes.  Kemball and Bishop (Ref.~\cite{Kemball-Bishop-91}) have
suggested and tested several modifications of this latter algorithm
to improve its convergence properties.

The final algorithm we discuss, and the main subject of this paper,
poses the apparent horizon equation $H(h) = 0$ as a nonlinear elliptic
(boundary-value) PDE for $h$ on the angular-coordinate space $\A$.
Finite differencing this PDE on an angular-coordinate grid
$\{ (\theta_\K,\phi_\K) \}$ gives a set of simultaneous nonlinear
algebraic equations for the unknown values $\{ h(\theta_\K,\phi_\K) \}$,
which are then solved by some variant of Newton's method.  This
\defn{Newton's-method} algorithm (we continue to use this term even
if a modification of Newton's method is actually used) has been used
in axisymmetric spacetimes by a number of authors, for example
Refs.~\cite{Eardley-75,Cook-PhD,Cook-York-90,Cook-Abrahams-92,%%%
Thornburg-PhD}, and is also applicable in fully general spacetimes
when the coordinates have a (locally) polar spherical topology.
Huq (Ref.~\cite{Huq-talk-93}) has extended this algorithm to fully
general spacetimes with Cartesian-topology coordinates and finite
differencing, and much of our discussion remains applicable to his
extension.

The Newton's-method algorithm has three main parts: the computation
of the discrete horizon function $\H(\h)$, the computation of the
discrete horizon function's Jacobian matrix $\Jac[\H(\h)]$, and the
solution of the simultaneous nonlinear algebraic equations $\H(\h) = 0$.
We now discuss these in more detail.
%
%%%%%%%%%%%%%%%%%%%%%%%%%%%%%%%%%%%%%%%%%%%%%%%%%%%%%%%%%%%%%%%%%%%%%%%%%%%%%%%
%
\section{Computing the Horizon Function}
\label{sect-horizon-function}

In this section we discuss the details of the computation of the
discrete horizon function $\H(\h)$.  More precisely, first fix an
angular-coordinate grid $\{ (\theta_\K,\phi_\K) \}$.  Then, given a
\defn{trial horizon surface} $r = h(\theta,\phi)$, which need not
actually be an apparent horizon, we define $\h(\theta,\phi)$ to be
the discretization of $h(\theta,\phi)$ to the angular-coordinate
grid, and we consider the computation of $\H(\h)$ on the discretized
trial horizon surface, i.e.~at the points
$\{ ( {r=\h(\theta_\K,\phi_\K)}, {\theta=\theta_\K}, {\phi=\phi_\K} ) \}$.

The apparent horizon equation \eqref{eqn-horizon} defines
$H \equiv \two\! H$ in terms of the field variables and their
spatial derivatives on the trial horizon surface.  However, these
are typically known only at the (3-dimensional) grid points of the
underlying $3+1$ code of which the horizon finder is a part.  We
therefore extend $\two\! H$ to some (3-dimensional) neighborhood
$\N$ of the trial horizon surface, i.e.~we define an extended
horizon function $\three\! H$ on $\N$,
\begin{eqnarray}
\three\! H
	& = &	\del_i n^i + K_{ij} n^i n^j - K				\\
	& = &	\partial_i n^i
		+ (\partial_i \ln \sqrt{g}) n^i
		+ K_{ij} n^i n^j
		- K
					\, \text{.}	% text punctuation
							\label{eqn-3H(n-u)}
									%%%\\
\end{eqnarray}
To compute $\two \H(\h)$ on the (discretized) trial horizon surface,
we first compute $\three \H(\h)$ on the underlying $3+1$ code's
(3-dimensional) grid points in $\N$, then radially interpolate these
$\three \H$ values to the trial-horizon-surface position to obtain
$\two \H(\h)$,
\begin{mathletters}
							\label{eqn-2H(3H)}
\begin{equation}
\two \H(\theta,\phi)
	= \interp \Bigl( \three \H(r,\theta,\phi), r=\h(\theta,\phi) \Bigr)
					\, \text{,}	% text punctuation
\end{equation}
or equivalently
\begin{equation}
\two \H_\I
	= \interp \Bigl( \three \H_{\langle r\I \rangle}, r=\h_\I \Bigr)
					\, \text{,}	% text punctuation
\end{equation}
\end{mathletters}
where $\I$ is an angular grid-point index and the $\langle r\I \rangle$
subscript denotes that the interpolation is done independently at each
angular coordinate along the radial coordinate line
$\{ \theta=\theta_\I, \phi=\phi_\I \}$.  In practice any reasonable
interpolation method should work well here: Refs.~\cite{Cook-York-90,%%%
Cook-Abrahams-92} report satisfactory results using a spline interpolant;
in this work we use a Lagrange (polynomial) interpolant centered on
the trial-horizon-surface position, also with satisfactory results.
Neglecting the interpolation error, we can also write~\eqref{eqn-2H(3H)}
in the form
\begin{equation}
\two \H(\theta,\phi) = \three \H(r=\h(\theta,\phi), \theta, \phi)
					\, \text{.}	% text punctuation
						\label{eqn-2H(3H)-no-interp}
\end{equation}

We consider two basic types of methods for computing the extended
horizon function $\three \H(\h)$:
\begin{itemize}
\item	A \defn{2-stage} computation method uses two sequential
	numerical finite differencing stages, first explicitly
	computing $\s_i$ and/or $\n^i$ by numerically finite
	differencing $\h$, then computing $\three \H$ by
	numerically finite differencing $\s_i$ or $\n^i$.
\item	A \defn{1-stage} computation method uses only a single
	numerical (2nd) finite differencing stage, computing
	$\three \H$ directly in terms of $\h$'s 1st~and 2nd
	angular derivatives.
\end{itemize}
Figure~\ref{fig-H(h)-methods} illustrates this.

To derive the detailed equations for these methods, we substitute
\eqrefs{eqn-s-d(h)} and \eqrefs{eqn-n-u(s-d)} into \eqref{eqn-3H(n-u)}:
\begin{eqnarray}
\three\! H
	& = &	\del_i n^i + K_{ij} n^i n^j - K
									\\
	& = &	\partial_i n^i
		+ (\partial_i \ln \sqrt{g}) n^i
		+ K_{ij} n^i n^j
		- K
									\\
	& = &	\partial_i \frac{g^{ij} s_j}{(g^{kl} s_k s_l)^{1/2}}
		+ (\partial_i \ln \sqrt{g})
		  \frac{g^{ij} s_j}{(g^{kl} s_k s_l)^{1/2}}
		+ \frac{K^{ij} s_i s_j}{g^{kl} s_k s_l}
		- K
									\\
	& = &	{}
		\frac{A}{D^{3/2}}
		+ \frac{B}{D^{1/2}}
		+ \frac{C}{D}
		- K
					\, \text{,}	% text punctuation
							\label{eqn-3H(ABCD)}
									%%%\\
\end{eqnarray}
where the subexpressions $A$, $B$, $C$, and $D$ are given by
\begin{mathletters}
							\label{eqn-ABCD(s-d)}
\begin{eqnarray}
A	& = &	{}
		- (g^{ik} s_k) (g^{jl} s_l) \partial_i s_j
		- \thalf (g^{ij} s_j) \Bigl[ (\partial_i g^{kl}) s_k s_l \Bigr]
									\\
B	& = &	(\partial_i g^{ij}) s_j
		+ g^{ij} \partial_i s_j
		+ (\partial_i \ln \sqrt{g}) (g^{ij} s_j)
									\\
C	& = &	K^{ij} s_i s_j
									\\
D	& = &	g^{ij} s_i s_j
					\, \text{,}	% text punctuation
									%%%\\
\end{eqnarray}
\end{mathletters}
i.e.
\begin{mathletters}
							\label{eqn-ABCD(h)}
\begin{eqnarray}
A 	& = &	(g^{ur} - g^{uw} \partial_w h)
		(g^{vr} - g^{vw} \partial_w h)
		\partial_{uv} h
							\nonumber	\\
	&   &	\quad
		{}
		- \thalf
		  (g^{ir} - g^{iu} \partial_u h)
		  \Bigl[
		  \partial_i g^{rr}
		  - 2 (\partial_i g^{ru}) \partial_u h
		  + (\partial_i g^{uv}) (\partial_u h) (\partial_v h)
		  \Bigr]
									\\
B	& = &	\Bigl[
		\partial_i g^{ir} - (\partial_i g^{iu}) \partial_u h
		\Bigr]
		- g^{uv} \partial_{uv} h
		+ (\partial_i \ln \sqrt{g}) (g^{ir} - g^{iu} \partial_u h)
									\\
C 	& = &	K^{rr}
		- 2 K^{ru} \partial_u h
		+ K^{uv} (\partial_u h) (\partial_v h)
									\\
D	& = &	g^{rr}
		- 2 g^{ru} \partial_u h
		+ g^{uv} (\partial_u h) (\partial_v h)
					\, \text{.}	% text punctuation
									%%%\\
\end{eqnarray}
\end{mathletters}

Comparing the 1-stage and 2-stage methods, the 2-stage methods'
equations are somewhat simpler, so these methods are somewhat easier
to implement and somewhat cheaper (faster) to compute.  However, for
a proper comparison the cost of computing the horizon function must
be considered in conjunction with the cost of computing the horizon
function's Jacobian.  Compared to the 1-stage method, the 2-stage
methods double the effective radius of the net $\H(\h)$ finite
differencing molecules, and thus have 2(4)~times as many nonzero
off-diagonal Jacobian elements for a 2(3)-dimensional code.  In practice
the cost of computing these extra Jacobian elements for the 2-stage
methods more than outweighs the slight cost saving in evaluating the
horizon function.  We discuss the relative costs of the different
methods further in section~\ref{sect-methods-comparison}.
%
%%%%%%%%%%%%%%%%%%%%%%%%%%%%%%%%%%%%%%%%%%%%%%%%%%%%%%%%%%%%%%%%%%%%%%%%%%%%%%%
%
\section{Computing the Jacobian}
\label{sect-Jacobian}

In this section we discuss the details of the computation of the
Jacobian matrix $\Jac[\H(\h)]$ of the horizon function $\H(\h)$ on
a given trial horizon surface.
%
%%%%%%%%%%%%%%%%%%%%%%%%%%%%%%%%%%%%%%%%
%
\subsection{Computing the Jacobian of a Generic Function $\P(\Q)$}

We consider first the case of a generic function $P(Q)$ in $d$~dimensions,
finite differenced using $N$-point molecules.  We define the Jacobian
matrix of the discrete $\P(\Q)$ function by
\begin{mathletters}
							\label{eqn-Jac[P(Q)]}
\begin{equation}
\Jac[\P(\Q)]_{\I\J} = \frac{\partial \P_\I}{\partial \Q_\J}
					\, \text{,}	% text punctuation
\end{equation}
or equivalently by the requirement that
\begin{equation}
\delta \P_\I \equiv
	\Bigl[ \P(\Q + \delta \Q) - \P(\Q) \Bigr]_\I
	= \Jac[\P(\Q)]_{\I\J} \cdot \delta \Q_\J
\end{equation}
\end{mathletters}
for any infinitesimal perturbation $\delta \Q$ of $\Q$.

We assume that $\P$ is actually a {\em local\/} grid function of $\Q$,
so the Jacobian matrix is sparse.  (For example, this would preclude
the nonlocal 4th~order \defn{compact differencing} methods described by
Refs.~\cite{Ciment-Leventhal-75,Hirsh-75}.)  We assume that by exploiting
the locality of the discrete $\P(\Q)$ function, any single $\P_\I$ can
be computed in $O(1)$ time, independent of the grid size.
%
%%%%%%%%%%%%%%%%%%%%
%
\subsubsection{Computing Jacobians by Numerical Perturbation}
\label{sect-numerical-perturbation}

We consider two general methods for computing the Jacobian matrix
$\Jac[\P(\Q)]$.  The first of these is the \defn{numerical perturbation}
method.  This involves numerically perturbing $\Q$ and examining the
resulting perturbation in $\P(\Q)$,
\begin{equation}
\Jac[\P(\Q)]_{\I\J} \approx
	\left[ \frac{\P(\Q + \mu \e^{(\J)}) - \P(\Q)}{\mu} \right]_\I
					\, \text{,}	% text punctuation
						\label{eqn-Jac[P(Q)]-NP}
\end{equation}
where $\e^{(\J)}$ is a Kronecker-delta vector defined by
\begin{equation}
\left[ \e^{(\J)} \right]_\I
	= \left\{
	  \begin{array}{l@{\quad}l}
	  1	& \text{if $\I = \J$}	\\
	  0	& \text{otherwise}	%%%\\
	  \end{array}
	  \right.
					\, \text{,}	% text punctuation
\end{equation}
and $\mu$ is a ``small'' perturbation amplitude.  This computation
of the Jacobian proceeds by columns: for each $\J$, $\Q_\J$ is perturbed,
and the resulting perturbation in $\P(\Q)$ gives the $\J$th column of
the Jacobian matrix.

The perturbation amplitude $\mu$ should be chosen to balance the
truncation error of the one-sided finite difference approximation
\eqref{eqn-Jac[P(Q)]-NP} against the numerical loss of significance
caused by subtracting the nearly equal quantities $\P(\Q + \mu \e^{(\J)})$
and $\P(\Q)$.  Refs.~\cite{Curtis-Reid-74,Stoer-Bulirsch-80-quote}
discuss the choice of $\mu$, and conclude that if $\P(\Q)$ can be
evaluated with an accuracy of $\varepsilon$, then
$\mu \approx \sqrt{\varepsilon}$ ``seems to work the best''.  In
practice the choice of $\mu$ isn't very critical for horizon finding.
Values of $10^{-4}$ to $10^{-6}$ seem to work well, and the inaccuracies
in the Jacobian matrix resulting from these values of $\mu$ don't
seem to be a significant problem.

This method of computing Jacobians requires no knowledge of the
$\P(\Q)$ function's internal structure.  In particular, the $\P(\Q)$
function may involve arbitrary nonlinear computations, including
multiple sequential stages of finite differencing and/or interpolation.
This method is thus directly applicable to the $\two \H(\h)$ computation.

Assuming that $\P(\Q)$ is already known, computing $\Jac[\P(\Q)]$ by
numerical perturbation requires a total of $N^d$ $\P_\I$ evaluations
at each grid point, i.e.~it requires a perturbed-$\P_\I$ evaluation
for each nonzero Jacobian element.
%
%%%%%%%%%%%%%%%%%%%%
%
\subsubsection{Computing Jacobians by Symbolic Differentiation}
\label{sect-symbolic-differentiation}

An alternate method of computing the Jacobian matrix $\Jac[\P(\Q)]$
is by \defn{symbolic differentiation}.  This method makes explicit
use of the finite differencing scheme used to compute the discrete
$\P(\Q)$ function.

Suppose first that the continuum $P(Q)$ function is a position-dependent
local {\em linear\/} differential operator, discretely approximated
by a position-dependent local finite difference molecule $\MM$,
\begin{equation}
\P_\I = \sum_{\M \in \MM(\I)} \MM(\I)_\M \Q_{\I+\M}
					\, \text{.}	% text punctuation
							\label{eqn-P(Q)-mol}
\end{equation}
Differentiating this, we have
\begin{equation}
\Jac[\P(\Q)]_{\I\J} \equiv \frac{\partial \P_\I}{\partial \Q_\J}
	= \left\{
	  \begin{array}{l@{\quad}l}
	  \MM(\I)_{\J-\I}	& \text{if $\J-\I \in \MM(\I)$}	\\
	  0			& \text{otherwise}		%%%\\
	  \end{array}
	  \right.
					\, \text{,}	% text punctuation
						\label{eqn-Jac[P(Q)]-from-mol}
\end{equation}
so that the molecule coefficients at each grid point give the
corresponding row of the Jacobian matrix.

More generally, suppose $P$ is a position-dependent local nonlinear
algebraic function of $Q$ and some finite number of $Q$'s derivatives,
say
\begin{equation}
P = P(Q, \partial_i Q, \partial_{ij} Q)
					\, \text{.}	% text punctuation
						\label{eqn-P(Q)-and-derivs}
\end{equation}

Logically, the Jacobian matrix $\Jac[\P(\Q)]$ is defined
(by \eqref{eqn-Jac[P(Q)]}) in terms of the linearization of the
discrete (finite differenced) $\P(\Q)$ function.  However, as
illustrated in figure~\ref{fig-linearize-vs-FD}, if the discretization
(the finite differencing scheme) commutes with the linearization,
we can instead compute the Jacobian by first linearizing the continuum
$P(Q)$ function, then finite differencing this (continuum) linearized
function.  (This method of computing the Jacobian is essentially just
the ``Jacobian part'' of the Newton-Kantorovich algorithm for solving
nonlinear elliptic PDEs.)

That is, we first linearize the continuum $P(Q)$ function,
\begin{eqnarray}
\delta P
	& = &	\frac{\partial P}{\partial Q}
		\, {\delta Q}
		+ \frac{\partial P}{\partial (\partial_i Q)}
		  \, \delta \partial_i Q
		+ \frac{\partial P}{\partial (\partial_{ij} Q)}
		  \, \delta \partial_{ij} Q				\\
	& = &	\frac{\partial P}{\partial Q}
		\, {\delta Q}
		+ \frac{\partial P}{\partial (\partial_i Q)}
		  \, \partial_i \delta Q
		+ \frac{\partial P}{\partial (\partial_{ij} Q)}
		  \, \partial_{ij} \delta Q
					\, \text{.}	% text punctuation
					\label{eqn-P(Q)-and-derivs-linearized}
									%%%\\
\end{eqnarray}
We then view the linearized function $\delta P(\delta Q)$ as a
linear differential operator, and discretely approximate it by
the position-dependent finite difference molecule
\begin{equation}
\MM = \frac{\partial P}{\partial Q} \II
      + \frac{\partial P}{\partial (\partial_i Q)} \dd_i
      + \frac{\partial P}{\partial (\partial_{ij} Q)} \dd_{ij}
					\, \text{,}	% text punctuation
					\label{eqn-mol-from-P(Q)-Jac-coeffs}
\end{equation}
where $\II$ is the identity molecule and $\dd_i$ and $\dd_{ij}$ are
finite difference molecules discretely approximating $\partial_i$
and $\partial_{ij}$ respectively.  Finally, we apply
\eqref{eqn-Jac[P(Q)]-from-mol} to the molecule $\MM$ defined by
\eqref{eqn-mol-from-P(Q)-Jac-coeffs} to obtain the desired Jacobian
matrix $\Jac[\P(\Q)]$.

In practice, there's no need to explicitly form the molecule $\MM$
-- the Jacobian matrix elements can easily be assembled directly from
the known $\II$, $\dd_i$, and $\dd_{ij}$ molecule coefficients and
the \defn{Jacobian coefficients} $\partial P / \partial Q$,
$\partial P / \partial (\partial_i Q)$, and
$\partial P / \partial (\partial_{ij} Q)$.  Once these coefficients
are known, the assembly of the actual Jacobian matrix elements is very
cheap, requiring only a few arithmetic operations per matrix element
to evaluate \eqrefs{eqn-mol-from-P(Q)-Jac-coeffs}
and~\eqrefb{eqn-Jac[P(Q)]-from-mol}.  The main cost of computing a
Jacobian matrix by symbolic differentiation is thus the computation
of the Jacobian coefficients themselves.  Depending on the functional
form of the $P(Q)$ function, there may be anywhere from 1 to 10
coefficients, although in practice these often have many common
subexpressions.

In other words, where the numerical perturbation method requires
a $\P_\I$ evaluation per nonzero Jacobian {\em element\/}, the
symbolic differentiation method requires the computation of ``a few''
Jacobian-coefficient subexpressions per Jacobian {\em row\/}.  More
precisely, suppose the computation of all the Jacobian coefficients
at a single grid point is $J$ times as costly as a $\P_\I$ evaluation.
Then the symbolic differentiation method is approximately $N^d/J$
times more efficient than the numerical perturbation method.
%
%%%%%%%%%%%%%%%%%%%%%%%%%%%%%%%%%%%%%%%%
%
\subsection{Semantics of the Horizon Function Jacobian}
\label{sect-Jacobian-semantics}

We now consider the detailed semantics of the horizon function
Jacobian.  We define the Jacobian of $\H(\h) \equiv \two \H(\h)$,
$\Jac[\H(\h)] \equiv \Jac[\two \H(\h)]$, by
\begin{mathletters}
							\label{eqn-Jac[2H(h)]}
\begin{equation}
\Jac[\two \H(\h)]_{\I\J} = \frac{d \, \two \H_\I}{d \h_\J}
					\, \text{,}	% text punctuation
\end{equation}
(where $\I$ and $\J$ are angular grid-point indices), or equivalently
by the requirement that
\begin{equation}
\delta \two \H_\I
	\equiv \Bigl[ \two \H(\h + \delta \h) - \two \H(\h) \Bigr]_\I
	= \Jac[\two \H(\h)]_{\I\J} \cdot \delta \h_\J
\end{equation}
\end{mathletters}
for any infinitesimal perturbation $\delta \h$.  Here $\I$ and $\J$
are both angular (2-dimensional) grid-point indices.  Notice that this
definition uses the {\em total\/} derivative $d \, \two \H / d \h$.
This is because $\two \H(\h)$ is defined to always be evaluated
{\em at the position $r = \h(\theta,\phi)$ of the trial horizon surface\/},
so the Jacobian $\Jac[\two \H(\h)]$ must take into account not only
the direct change in $\two \H$ at a fixed position due to a perturbation
in $\h$, but also the implicit change in $\two \H$ caused by the
field-variable coefficients in $\two \H$ being evaluated at a
perturbed position $r = \h(\theta,\phi)$.

It's also useful to consider the Jacobian $\Jac[\three \H(\h)]$ of
the extended horizon function $\three \H(\h)$, which we define
analogously by
\begin{mathletters}
							\label{eqn-Jac[3H(h)]}
\begin{equation}
\Jac[\three \H(\h)]_{\I\J} = \frac{\partial \, \three \H_\I}{\partial \h_\J}
					\, \text{,}	% text punctuation
\end{equation}
or equivalently by the requirement that
\begin{equation}
\delta \, \three \H_\I
	\equiv
	\Bigl[ \three \H(\h + \delta \h) - \three \H(\h) \Bigr]_\I
	= \Jac[\three \H(\h)]_{\I\J} \cdot \delta \h_\J
\end{equation}
\end{mathletters}
for any infinitesimal perturbation $\delta \h$.  Here $\I$ is a
3-dimensional grid-point index for $\three \H$, while $\J$ is an
(angular) 2-dimensional grid-point index for $\h$.  In contrast
to $\Jac[\two \H(\h)]$, this definition uses the {\em partial\/}
derivative $\partial \, \three \H / \partial \h$.  This is because
we take $\three \H(\h)$ to be evaluated at a fixed position (a
grid point in the neighborhood $\N$ of the trial horizon surface)
{\em which doesn't change with perturbations in $\h$\/}, so
$\Jac[\three \H(\h)]$ need only take into account the direct
change in $\three \H$ at a fixed position due to a perturbation
in $\h$.

$\Jac[\three \H(\h)]$ thus has much simpler semantics than
$\Jac[\two \H(\h)]$.  We have found $\Jac[\three \H(\h)]$ very
useful, both as an intermediate variable in the computation of
$\Jac[\two \H(\h)]$ (described in the next section), and also
conceptually, as an aid to {\em thinking\/} about the Jacobians.
%
%%%%%%%%%%%%%%%%%%%%%%%%%%%%%%%%%%%%%%%%
%
\subsection{Computing the Horizon Function Jacobian}
\label{sect-computing-horizon-function-Jacobian}

Table~\ref{tab-methods-comparison} (discussed further in
section~\ref{sect-methods-comparison}) summarizes all the
Jacobian-computation methods in this paper, which we now describe
in detail.  We tag each method with a shorthand \defn{code}, which
gives the method's basic properties: whether it computes
$\Jac[\two \H(\h)]$ directly or computes $\Jac[\three \H(\h)]$ as
an intermediate step, whether it uses symbolic differentiation or
numerical perturbation, and whether it uses a 1-stage or a 2-stage
horizon function computation.

The simplest methods for computing $\Jac[\two \H(\h)]$ are the
\defn{2-dimensional} ones, which work directly with $\two \H(\h)$
in angular-coordinate space, without computing $\Jac[\three \H(\h)]$
as an intermediate step.  Since $\two \H(\h)$ isn't given by a
simple molecule operation of the form \eqref{eqn-P(Q)-mol}, symbolic
differentiation isn't directly applicable here.  However, numerical
perturbation in angular-coordinate space is applicable, using either
a 1-stage or a 2-stage method to compute $\two \H(\h)$.  We refer
to the resulting Jacobian computation methods as the \defn{2d.np.1s}
and \defn{2d.np.2s} methods respectively.

Our remaining methods for computing $\Jac[\two \H(\h)]$ are
all \defn{3-dimensional} ones, which first explicitly compute
$\Jac[\three \H(\h)]$, then compute $\Jac[\two \H(\h)]$ from this
in the manner described below.

If $\three \H(\h)$ is computed using the 1-stage method, i.e.~via
\eqrefs{eqn-3H(ABCD)} and~\eqrefb{eqn-ABCD(h)}, then either numerical
perturbation or symbolic differentiation may be used to compute
$\Jac[\three \H(\h)]$.  We refer to these as the \defn{3d.np.1s}
and \defn{3d.sd.1s} methods respectively.  The symbolic-differentiation
Jacobian coefficients for the 3d.sd.1s method are tabulated in
appendix~\ref{app-SD-Jac-coeffs}.

Alternatively, if $\three \H(\h)$ is computed using a 2-stage
method, then $\Jac[\three \H(\h)]$ may be computed either by the
simple numerical perturbation of $\three \H(\h)$ (the \defn{3d.np.2s}
method), or by separately computing the Jacobians of the individual
stages and matrix-multiplying them together.  For the latter case,
either numerical perturbation or symbolic differentiation may be
used to compute the individual-stage Jacobians, giving the
\defn{3d.np2.2s} and \defn{3d.sd2.2s} methods respectively.  The
symbolic-differentiation Jacobian coefficients for the 3d.sd2.2s
method are tabulated in appendix~\ref{app-SD-Jac-coeffs}.

For any of the 3-dimensional methods, once $\Jac[\three \H(\h)]$
is known, we compute $\Jac[\two \H(\h)]$ as follows:
\begin{eqnarray}
\Jac[\two \H(\h)]_{\I\J}
	& \equiv &
		\frac{d \, \two \H_\I}{d \h_\J}
									\\
	& = &	\frac{d \, \two \H(\theta_\I, \phi_\I)}
		     {d \h(\theta_\J, \phi_\J)}
									\\
	& = &	\frac{d \,
		      \three \H(r=\h(\theta_\I,\phi_\I), \theta_\I, \phi_\I)}
		     {d \h(\theta_\J, \phi_\J)}
		\qquad
		\text{(by~\eqref{eqn-2H(3H)-no-interp})}
									\\
	& = &	\left.
		\frac{\partial \, \three \H(r, \theta_\I, \phi_\I)}
		     {\partial \h(\theta_\J, \phi_\J)}
		\right|_{r = \h(\theta_\I, \phi_\I)}
		+
		\left.
		\frac{\partial \, \three \H(r, \theta_\I, \phi_\I)}
		     {r}
		\right|_{r = \h(\theta_\I, \phi_\I)}
									\\
	& = &	\interp \Bigl(
			\Jac[\three \H(\h)]_{{\langle r\I \rangle} \J},
			r = \h_\I
			\Bigr)
		+
		\interp' \Bigl(
			 \three \H_{\langle r\I \rangle}, r = \h_\I
			 \Bigr)
					\, \text{,}	% text punctuation
					\label{eqn-Jac[2H(h)]-from-Jac[3H(h)]}
									%%%\\
\end{eqnarray}
where the $\langle r\I \rangle$ subscripts in
\eqref{eqn-Jac[2H(h)]-from-Jac[3H(h)]} denote that the interpolations
are done along the radial line $\{ \theta=\theta_\I, \phi=\phi_\I \}$,
analogously to \eqref{eqn-2H(3H)}, and where we neglect the
interpolation errors in~\eqref{eqn-Jac[2H(h)]-from-Jac[3H(h)]}.

Notice that the \,\,$\interp'(\dots)$\,\, term
in~\eqref{eqn-Jac[2H(h)]-from-Jac[3H(h)]} may be computed very cheaply
using the same $\three \H$ data values used in computing $\two \H$,
cf.~\eqref{eqn-2H(3H)}.  (The number of $\three \H$ data points used
in the radial interpolation at each angular grid position will probably
have to be increased by one to retain the same order of accuracy in the
\,\,$\interp'(\dots)$\,\, term in \eqref{eqn-Jac[2H(h)]-from-Jac[3H(h)]}
as in the \,\,$\interp(\dots)$\,\, term.)  It's thus easy to compute
$\Jac[\two \H(\h)]$ once $\Jac[\three \H(\h)]$ is known.
%
%%%%%%%%%%%%%%%%%%%%%%%%%%%%%%%%%%%%%%%%
%
\subsection{Comparing the Methods}
\label{sect-methods-comparison}

Table~\ref{tab-methods-comparison} summarizes all the
horizon-function and Jacobian computation methods described
in sections~\ref{sect-horizon-function}
and~\ref{sect-computing-horizon-function-Jacobian}.  The table also
shows which Jacobian matrices the methods use, the methods' measured
relative CPU times in our axisymmetric-spacetime (2-dimensional) code
(discussed further in appendix~\ref{app-code-details}), and our
estimates of the methods' approximate implementation effort
(programming complexity).

As can be seen from the table, for our implementation the 3d.sd.1s
method is by far the most efficient of the Jacobian computation
methods, being about a factor of $5$ faster than any of the numerical
perturbation methods.  In fact, the computation of the Jacobian
$\Jac[\two \H(\h)]$ by the 3d.sd.1s method is only $1.5$--$2$ times
more expensive than the simple evaluation of the horizon function
$\two \H(\h)$.

The relative performance of the different methods will of course
vary considerably from one implementation to another, and especially
between axisymmetric-spacetime (2-dimensional) and fully-general-spacetime
(3-dimensional) codes.  However, counting the number of operations
needed for each method shows that the 3d.sd.1s method should remain
the fastest for any reasonable implementation.  (We omit details of
the counting in view of their length and lack of general interest.)
Notably, the 3d.sd.1s method's relative advantage over the other
methods should be approximately a factor of the molecule diameter
{\em larger\/} for fully-general-spacetime (3-dimensional) codes
than for axisymmetric-spacetime (2-dimensional) codes such as ours.

Considering now the implementation efforts required by the various
methods, in general we find that these depend more on which Jacobian
matrices are involved than on how the Jacobians are computed:  The
2-dimensional methods, involving only $\Jac[\two \H(\h)]$, are the
easiest to implement, while the 3-dimensional methods involving (only)
$\Jac[\two \H(\h)]$ and $\Jac[\three \H(\h)]$ are somewhat harder to
implement.  The 3-dimensional methods involving the individual-stage
Jacobians $\Jac[\s_i(\h)]$, $\Jac[\n^i(\h)]$, $\Jac[\three \H(\s_i)]$,
and/or $\Jac[\three \H(\n^i)]$ are considerably more difficult to
implement, due to these Jacobians' more complicated sparsity patterns.

All the Jacobian matrices are highly sparse, and for reasonable
efficiency it's essential to exploit this sparsity in their storage
and computation.  We have done this in our code, and our CPU-time
measurements and implementation-effort estimates all reflect this.
We briefly describe our sparse-Jacobian storage scheme in
appendix~\ref{app-sparse-Jacobian-storage}.  This scheme is very
efficient, but its programming is a significant fraction of the
overall Jacobian implementation effort, especially for the
individual-stage Jacobians.

Comparing numerical perturbation and symbolic differentiation methods,
we had previously suggested (Ref.~\cite{Thornburg-PhD-SD-Jac-comments})
that symbolic-differentiation Jacobian computations would be very
difficult to implement, necessarily requiring substantial support
from a (computer) symbolic computation system.  Several colleagues
have expressed similar opinions to us.  We had also previous suggested
(Ref.~\cite{Thornburg-PhD-SD-Jac-comments}) that due to the structure
of the $H(h)$ function, a Jacobian-coefficient formalism of the type
described in sections~\ref{sect-symbolic-differentiation}
and~\ref{sect-computing-horizon-function-Jacobian} would not be
valid for the horizon function, so symbolic differentiation methods
would require explicitly differentiating the finite difference
equations.

These suggestions have proven to be incorrect: using the
Jacobian-coefficient formalism described in
sections~\ref{sect-symbolic-differentiation}
and~\ref{sect-computing-horizon-function-Jacobian}, only the continuum
equations need be differentiated, and this is easily done by hand.
More generally, using this formalism we find the actual programming
of the symbolic differentiation methods to be only moderately more
difficult than that of the numerical perturbation methods.  Some of
the Jacobian coefficients tabulated in appendix~\ref{app-SD-Jac-coeffs}
are fairly complicated, but no more so than many other computations
in $3+1$ numerical relativity.

In order to be confident of the correctness of any of the
Jacobian-computation methods except the simple 2-dimensional
numerical perturbation ones, we feel that it's highly desirable to
program an independent method (which may be programmed for simplicity
at the expense of efficiency) and make an end-to-end comparison of
the resulting Jacobian matrices.  (We have successfully done this
for each of the Jacobian matrices computed by each of the methods
listed in table~\ref{tab-methods-comparison}, and our implementation-effort
estimates there include doing this.)  If, and only if, the Jacobians
agree to within the expected truncation error of the numerical-perturbation
Jacobian approximation~\eqref{eqn-Jac[P(Q)]-NP}, then we can have a
high degree of confidence that both calculations are correct.  If
they disagree, then we find the detailed pattern of which matrix
elements differ to be a very useful debugging aid.

Summarizing our comparisons, then, we find that the best Jacobian
computation method is clearly the 3d.sd.1s one.  It's much more
efficient than any of the other methods, and still quite easy to
implement.
%
%%%%%%%%%%%%%%%%%%%%%%%%%%%%%%%%%%%%%%%%%%%%%%%%%%%%%%%%%%%%%%%%%%%%%%%%%%%%%%%
%
\section{Convergence Tests}
\label{sect-convergence-tests}

Before continuing our discussion of Newton's-method horizon finding,
in this section we digress to consider the convergence of finite
differencing computations to the continuum limit.

As has been forcefully emphasized by Choptuik
(Refs.~\cite{Choptuik-PhD,Choptuik-91,Choptuik-Goldwirth-Piran-92}),
a careful comparison of a finite differencing code's numerical results
at different grid resolutions can yield very stringent tests of the
code's numerical performance and correctness.  In particular, such
\defn{convergence tests} can yield reliable numerical estimates of a
code's {\em external\/} errors, i.e.~of the deviation of the code's
results from those that would be obtained by exactly solving the
continuum equations.  With, and only with, such estimates available,
we can safely draw inferences about solutions of the continuum equations
from the code's (finite-resolution) numerical results.

To apply this technique in the horizon-finding context, suppose first
that the (a) true (continuum) apparent horizon position $h^\ast$ is
known.  For a convergence test in this case, we run the horizon finder
twice, using a 1:2 ratio of grid resolutions.  As discussed in detail
by Ref.~\cite{Choptuik-91}, if the code's numerical errors are dominated
by truncation errors from $n$th~order finite differencing, the numerically
computed horizon positions $\h$ must satisfy
\begin{mathletters}
							\label{eqn-h(dx:dx/2)}
\begin{eqnarray}
\h[\Delta x]
	& = &	h^\ast + (\Delta x)^n f + O((\Delta x)^{n+2})
						\label{eqn-h(dx:dx/2)-dx}
									\\
\h[\Delta x / 2]
	& = &	h^\ast + (\Delta x / 2)^n f + O((\Delta x)^{n+2})
						\label{eqn-h(dx:dx/2)-dx/2}
									%%%\\
\end{eqnarray}
\end{mathletters}
at each grid point, where $\h[\Delta x]$ denotes the numerically
computed horizon position using grid resolution $\Delta x$, and $f$
is an $O(1)$ smooth function depending on various high order derivatives
of $h^\ast$ and the field variables, but {\em not\/} on the grid
resolution.  (We're assuming centered finite differencing here in
writing the higher order terms as $O((\Delta x)^{n+2})$, otherwise
they would only be $O((\Delta x)^{n+1})$.)  Neglecting the higher order
terms, i.e.~in the limit of small $\Delta x$, we can eliminate $f$ to
obtain a direct relationship between the code's errors at the two
resolutions,
\begin{equation}
\frac{\h[\Delta x / 2] - h^\ast}{\h[\Delta x] - h^\ast}
	= \frac{1}{2^n}
					\, \text{,}	% text punctuation
						\label{eqn-conv-test-dx:dx/2}
\end{equation}
which must hold at each grid point common to the two grids.

To test how well any particular set of (finite-resolution) numerical
results satisfies this convergence criterion, we plot a scatterplot
of the high-resolution errors $\h[\Delta x / 2] - h^\ast$ against the
low-resolution errors $\h[\Delta x] - h^\ast$ at the grid points common
to the two grids.  If, and given the arguments of Ref.~\cite{Choptuik-91},
in practice {\em only\/} if, the error expansions~\eqref{eqn-h(dx:dx/2)}
are valid with the higher order error terms negligible, i.e.~if and only
if the errors are indeed dominated by the expected $n$th~order finite
difference truncation errors, then all the points in the scatterplot
will fall on a line through the origin with slope $1/2^n$.

Now suppose the true (continuum) apparent horizon position $h^\ast$ is
unknown.  For a convergence test in this case, we run the horizon finder
three times, using a 1:2:4 ratio of grid resolutions.  Analogously to
the 2-grid case, we now have
\begin{mathletters}
						\label{eqn-h(dx:dx/2:dx/4)}
\begin{eqnarray}
\h[\Delta x]
	& = &	h^\ast + (\Delta x)^n f + O((\Delta x)^{n+2})
									\\
\h[\Delta x / 2]
	& = &	h^\ast + (\Delta x / 2)^n f + O((\Delta x)^{n+2})
									\\
\h[\Delta x / 4]
	& = &	h^\ast + (\Delta x / 4)^n f + O((\Delta x)^{n+2})
					\, \text{,}	% text punctuation
									%%%\\
\end{eqnarray}
\end{mathletters}
at each grid point, with $f$ again independent of the grid resolution.
Again neglecting the higher order terms, we can eliminate both $f$
and $h^\ast$ to obtain the \defn{3-grid} convergence criterion
\begin{equation}
\frac{\h[\Delta x / 2] - \h[\Delta x / 4]}{\h[\Delta x] - \h[\Delta x / 2]}
	= \frac{1}{2^n}
					\label{eqn-conv-test-dx:dx/2:dx/4}
\end{equation}
which must hold at each grid point common to the three grids.  We
test this criterion using a scatterplot technique analogous to that
for the 2-grid criterion~\eqref{eqn-conv-test-dx:dx/2}.

We emphasize that for a 3-grid convergence test of this type, the
true continuum solution $h^\ast$ need not be known.  In fact, nothing
in the derivation actually requires $h^\ast$ to be the true continuum
horizon position -- it need only be the true continuum solution to
some continuum equation such that the truncation error formulas
\eqref{eqn-h(dx:dx/2:dx/4)} hold.  We make use of this latter case
in sections~\ref{sect-Newton-Kantorovich-method}
and~\ref{sect-global-conv-HSF-errors} to apply 3-grid convergence
tests to intermediate Newton iterates (trial horizon surfaces) of
our horizon finder.

For both the 2-grid and the 3-grid convergence test, we find that
the {\em pointwise\/} nature of the scatterplot comparison makes it
significantly more useful than a simple comparison of gridwise norms.
In particular, the scatterplot comparison clearly shows convergence
problems which may occur only in a small subset of the grid points
(for example near a boundary), which would be ``washed out'' in a
comparison of gridwise norms.

Notice also that the parameter $n$, the order of the convergence, is
(should be) known in advance from the form of the finite differencing
scheme.  Thus the slope-$1/2^n$ line with which the scatterplot points
are compared isn't fitted to the data points, but is rather an a~priori
prediction with {\em no\/} adjustable parameters.  Convergence tests
of this type are thus a very strong test of the validity of the finite
differencing scheme and the error expansions \eqref{eqn-h(dx:dx/2)}
or~\eqref{eqn-h(dx:dx/2:dx/4)}.
%
%%%%%%%%%%%%%%%%%%%%%%%%%%%%%%%%%%%%%%%%%%%%%%%%%%%%%%%%%%%%%%%%%%%%%%%%%%%%%%%
%
\section{Solving the Nonlinear Algebraic Equations}
\label{sect-nonlinear-algebraic-equations}

Returning to our specific discussion of horizon finding, we now
discuss the details of using Newton's method or a variant to solve
the simultaneous nonlinear algebraic equations $\H(\h) = 0$.
%
%%%%%%%%%%%%%%%%%%%%%%%%%%%%%%%%%%%%%%%%
%
\subsection{Newton's Method}
\label{sect-Newton's-method}

The basic Newton's-method algorithm is well known:  At each iteration,
we first linearize the discrete $\H(\h)$ function about the current
approximate solution $\h^{(k)}$,
\begin{equation}
\H(\h^{(k)} + \delta \h)
	= \H(\h^{(k)})
	  + \Jac[\H(\h^{(k)})] \cdot \delta \h
	  + O(\|\delta \h\|^2)
					\, \text{,}	% text punctuation
						\label{eqn-H(h)-linearized}
\end{equation}
where $\delta \h$ now denotes a finite perturbation in $\h$, and
where $\Jac[\H(\h^{(k)})]$ denotes the Jacobian matrix $\Jac[\H(\h)]$
evaluated at the point $\h = \h^{(k)}$.  We then neglect the higher
order (nonlinear) terms and solve for the perturbation $\delta \h^{(k)}$
such that $\H(\h^{(k)} + \delta \h^{(k)}) = 0$.  This gives the
simultaneous linear algebraic equations
\begin{equation}
\Jac[\H(\h^{(k)})] \cdot \delta \h^{(k)} = - \H(\h^{(k)})
						\label{eqn-Newton-delta-h}
\end{equation}
to be solved for $\delta \h^{(k)}$.  Finally, we update the approximate
solution via
\begin{equation}
\h^{(k+1)} \leftarrow \h^{(k)} + \delta \h^{(k)}
					\, \text{,}	% text punctuation
						\label{eqn-Newton-h-update}
\end{equation}
and repeat the iteration until some convergence criterion is satisfied.

Notice that here we're using the word ``convergence'' in a very
different sense from that of section~\ref{sect-convergence-tests}
-- here it refers to the \defn{iteration-convergence} of the Newton
iterates $\h^{(k)}$ to the exact solution $\h^\ast$ of the discrete
equations, whereas there it refers to the
\defn{finite-difference-convergence} of a finite difference
computation result $\h[\Delta x]$ to its continuum limit $h^\ast$
as the grid resolution is increased.

Once the current solution estimate $\h^{(k)}$ is reasonably close
to $\h^\ast$, i.e.~in practice once the trial horizon surface is
reasonably close to the (an) apparent horizon, Newton's method converges
extremely rapidly.  In particular, once the linear approximation in
\eqref{eqn-H(h)-linearized} is approximately valid, Newton's method
roughly squares the relative error $\|\h - \h^\ast\| / \|\h^\ast\|$
at each iteration, and can thus bring the error down to a negligible
value in only a few (more) iterations.  (This rapid \defn{quadratic}
convergence depends critically on the mutual consistency of the horizon
function and Jacobian matrix used in the computation, and is thus a
useful diagnostic for monitoring the Jacobian's correctness.)  (For a
detailed discussion of Newton's method, including precise formulations
and proofs of these statements, see, for example,
Ref.~\cite{Stoer-Bulirsch-80-Newton's-method}.)

However, if the initial guess $\h^{(0)}$ for the horizon position, or
more generally any Newton iterate (trial horizon surface) $\h^{(k)}$,
differs sufficiently from $\h^\ast$ so that the linear approximation
in \eqref{eqn-H(h)-linearized} isn't approximately valid, then Newton's
method may converge poorly, or fail to converge at all.
%
%%%%%%%%%%%%%%%%%%%%%%%%%%%%%%%%%%%%%%%%
%
\subsection{Modifications of Newton's Method}
\label{sect-modifications-of-Newton's-method}

Unfortunately, as discussed in section~\ref{sect-global-conv-HSF-errors},
for certain types of initial guesses Newton's method fails to converge
unless the initial guess is very close to the exact solution of the
finite difference equations.  There's an extensive numerical analysis
literature on more robust \defn{modified Newton} algorithms for solving
nonlinear algebraic equations, for example Refs.~\cite{Bank-Rose-80,%%%
Bank-Rose-81,Dennis-Schnabel-83,MINPACK,Numerical-Recipes-2nd-edition,%%%
ZIB-90-11,ZIB-91-10}.  We have found Ref.~\cite{Dennis-Schnabel-83} to
be a particularly useful introduction to this topic.

For horizon finding, the Jacobian matrix's size is the number of
angular grid points on the horizon surface.  This is generally large
enough that it's important for the nonlinear-algebraic-equations
solver to support treating the Jacobian as either a band matrix
(for axisymmetric-spacetime codes) or a fully general sparse matrix
(for fully-general-spacetime codes).  It's also desirable for the
nonlinear-algebraic-equation solver to permit explicit bounds on the
solution vector, so as to ensure the trial horizon surfaces never
fall outside the radial extent of the code's main 3-dimensional grid.
Unfortunately, these requirements rule out many nonlinear-algebraic-equation
software packages.

For the sake of expediency, in the present work we chose to write
our own implementation of a relatively simple modified-Newton
algorithm, the \defn{line search} algorithm described by
Refs.~\cite{Dennis-Schnabel-83,Numerical-Recipes-2nd-edition}.
However, a much better long-term solution would be to use an extant
nonlinear-algebraic-equations code embodying high-quality implementations
of more sophisticated algorithms, such as the {\sc GIANT} code described
by Refs.~\cite{ZIB-90-11,ZIB-91-10}.  We would expect Newton's-method
horizon-finding codes using such software to be considerably more
robust and efficient than our present code.

The modified-Newton algorithm used in this work, the line-search
algorithm of Refs.~\cite{Dennis-Schnabel-83,Numerical-Recipes-2nd-edition},
is identical to the basic Newton's-method algorithm, except that
the Newton's-method update \eqref{eqn-Newton-h-update} is modified
to $\h^{(k+1)} \leftarrow \h^{(k)} + \lambda \, \delta \h^{(k)}$,
where $\lambda \in (0,1]$ is chosen at each \defn{outer} iteration
by an inner \defn{line search} iteration to ensure that $\|\H\|_2$
decreases monotonically.  Refs.~\cite{Dennis-Schnabel-83,%%%
Numerical-Recipes-2nd-edition} show that such a choice of $\lambda$
is always possible, and describe an efficient algorithm for it.
Sufficiently close to the solution $\h^\ast$, this algorithm always
chooses $\lambda = 1$, and so takes the same steps as Newton's method.
The overall modified-Newton algorithm thus retains the extremely
rapid convergence of Newton's method once the linear approximation
in \eqref{eqn-H(h)-linearized} is good.

The line-search algorithm described by Refs.~\cite{Dennis-Schnabel-83,%%%
Numerical-Recipes-2nd-edition} always begins by trying the basic Newton
step $\lambda = 1$.  For horizon finding, we have slightly modified
the algorithm to decrease the starting value of $\lambda$ if necessary
to ensure that $\h^{(k)} + \lambda \, \delta \h^{(k)}$ lies within
the radial extent of our code's main (3-dimensional) numerical grid
at each angular grid coordinate.  Our implementation of the algorithm
also enforces an upper bound (typically 10\%) on the relative change
$\|\lambda \, \delta \h^{(k)} / \h^{(k)}\|_\infty$ in any component
of $\h^{(k)}$ in a single outer iteration.  However, it's not clear
whether or not this latter restriction is a good idea: although it
makes the algorithm more robust when the $\H(\h)$ function is highly
nonlinear, it may slow the algorithm's convergence when the $\H(\h)$
function is only weakly nonlinear and the error in the initial guess
is large.  We give an example of this latter behavior in
section~\ref{sect-accuracy}.
%
%%%%%%%%%%%%%%%%%%%%%%%%%%%%%%%%%%%%%%%%
%
\subsection{The Newton-Kantorovich Method}
\label{sect-Newton-Kantorovich-method}

We have described the Newton's-method algorithm, and the more robust
modified versions of it, in terms of solving the discrete $\H(\h) = 0$
equations.  However, these algorithms can also be interpreted
directly in terms of solving the continuum $H(h) = 0$ equations.
This \defn{Newton-Kantorovich} method, and its relationship to the
discrete Newton's method, are discussed in detail by Ref.~\cite{Boyd}.

For the Newton-Kantorovich algorithm, at each iteration, we first
linearize the continuum differential operator $H(h)$ about the current
continuum approximate solution $h^{(k)}$,
\begin{equation}
H(h^{(k)} + \delta h)
	= H(h^{(k)}) + \Jacc[H(h^{(k)})] (\delta h) + O(\|\delta h\|^2)
					\, \text{,}	% text punctuation
					\label{eqn-continuum-H(h)-linearized}
\end{equation}
where $\delta h$ is now a finite perturbation in $h$, and where the
linear differential operator $\Jacc[H(h^{(k)})]$ is now the linearization
of the differential operator $H(h)$ about the point $h = h^{(k)}$.
We then neglect the higher order (nonlinear) terms and solve for the
perturbation $\delta h^{(k)}$ such that $H(h^{(k)} + \delta h^{(k)}) = 0$.
This gives the linear differential equation
\begin{equation}
\Jacc[H(h^{(k)})](\delta h^{(k)}) = - H(h^{(k)})
					\label{eqn-Newton-Kantorovich-dh}
\end{equation}
to be solved for $\delta h^{(k)}$.  Finally, we update the approximate
solution via
\begin{equation}
h^{(k+1)} \leftarrow h^{(k)} + \delta h^{(k)}
					\, \text{,}	% text punctuation
\end{equation}
and repeat the iteration until some convergence criterion is satisfied.

Now suppose we discretely approximate this continuum Newton-Kantorovich
algorithm by finite differencing the iteration equation
\eqref{eqn-Newton-Kantorovich-dh}.  If the finite differencing and
the linearization commute in the manner discussed in
section~\ref{sect-symbolic-differentiation}, then {\em this
finite-difference approximation to the Newton-Kantorovich algorithm
is in fact identical to the discrete Newton's-method algorithm applied
to the (discrete) $\H(\h) = 0$ equations obtained by finite differencing
the continuum $H(h) = 0$ equation\/}.  (In a simpler context, our
Jacobian-coefficient formalism described in
section~\ref{sect-symbolic-differentiation} essentially just exploits
the ``Jacobian part'' of this identity.)

Therefore, when using the discrete Newton's method to solve the
$\H(\h) = 0$ equations, we can equivalently view each Newton iterate
(trial horizon surface) $\h^{(k)}[\Delta x]$ as being a finite difference
approximation to the corresponding continuum Newton-Kantorovich iterate
(trial horizon surface) $h^{(k)}$.  As the grid resolution is increased,
each Newton iterate $\h^{(k)}[\Delta x]$ should therefore show proper
finite-difference-convergence {\em regardless of the iteration-convergence
or iteration-divergence of the Newton or Newton-Kantorovich iteration
itself\/}.

Moreover, once we verify the individual Newton iterates'
finite-differencing-convergence (with a 3-grid convergence test), we
can safely extrapolate the iteration-convergence or iteration-divergence
of this discrete iteration to that of the continuum Newton-Kantorovich
algorithm applied to the (continuum) $H(h) = 0$ equations.  In other
words, by this procedure we can ascribe the iteration-convergence or
iteration-divergence of Newton's method to inherent properties of the
continuum $H(h) = 0$ equations, as opposed to (say) a finite differencing
artifact.  We make use of this in section~\ref{sect-global-conv-HSF-errors}.
%
%%%%%%%%%%%%%%%%%%%%%%%%%%%%%%%%%%%%%%%%%%%%%%%%%%%%%%%%%%%%%%%%%%%%%%%%%%%%%%%
%
\section{Global Convergence of the Horizon Finder}
\label{sect-global-convergence}

We now consider the global convergence behavior of the Newton's-method
horizon finding algorithm.  That is, how close must the initial guess
$\h^{(0)}$ be to the (an) exact solution $\h^\ast$ of the finite
difference equations in order for the iterates (trial horizon surfaces)
$\h^{(k)}$ to converge to $\h^\ast$?  In other words, how large is the
algorithm's radius of convergence?
%
%%%%%%%%%%%%%%%%%%%%%%%%%%%%%%%%%%%%%%%%
%
\subsection{Global Convergence for Schwarzschild Spacetime}

To gain a general picture of the qualitative behavior of $H(h)$ and
its implications for Newton's-method horizon finding, it's useful to
consider Schwarzschild spacetime.  We use the Eddington-Finkelstein
slicing, where the time coordinate is defined by requiring $t + r$
to be an ingoing null coordinate.  (These slices aren't maximal: $K$
is nonzero and spatially variable throughout the slices.)

Taking the black hole to be of dimensionless unit mass, the (only)
apparent horizon in such a slice is the coordinate sphere $r = 2$.
More generally, a straightforward calculation gives
\begin{equation}
H = \frac{2 (r-2)}{r^{3/2} \sqrt{r + 2}}
\end{equation}
for spherical trial horizon surfaces with coordinate radius $r$.
Figure~\ref{fig-Schw-H} shows $H(r)$ for these surfaces.  As expected,
$H = 0$ for the horizon $r = 2$.  However, notice that $H$ reaches a
maximum value at $r = r^{\max} = \thalf (3 + \sqrt{33}) \approx 4.372$,
and in particular that for $r > r^{\max}$, $H > 0$ and $dH / dr < 0$.
Because of this, almost any algorithm -- including Newton's method
and its variants -- which tries to solve $H(r) = 0$ using only local
information about $H(r)$, and which maintains the spherical symmetry,
will diverge towards infinity when started from within this region,
or if any intermediate iterate (trial horizon surface) ever enters it.

In fact, we expect broadly similar behavior for $H$ in any black hole
spacetime:  Given an asymptotically flat slice containing an apparent
horizon or horizons, consider any 1-parameter family of topologically
2-spherical nested trial horizon surfaces starting at the outermost
apparent horizon and extending outward towards the 2-sphere at spatial
infinity.  $H = 0$ for the horizon and for the 2-sphere at spatial
infinity, so $\|H\|$ must attain a maximum for some finite trial horizon
surface somewhere between these two surfaces.  We thus expect the
same general behavior as in the Schwarzschild-slice case, i.e.~divergence
to infinity if the initial guess or any intermediate iterate (trial
horizon surface) lies outside the maximum-$\|H\|$ surface.  This
argument isn't completely rigorous, since the algorithm could move
inward in an angularly anisotropic manner, but this seems unlikely.

Fortunately, in practice this isn't a problem: the black hole area
theorem places an upper bound on the size of an apparent horizon,
and this lets us avoid overly-large initial guesses, or restart the
Newton iteration if any intermediate iterate (trial horizon surface)
is too large.
%
%%%%%%%%%%%%%%%%%%%%%%%%%%%%%%%%%%%%%%%%
%
\subsection{Global Convergence in the Presence of
	    High-Spatial-Frequency Errors}
\label{sect-global-conv-HSF-errors}

Assuming the initial guess is close enough to the horizon for the
divergence-to-infinity phenomenon not to occur, we find the global
convergence behavior of Newton's method to depend critically on the
angular spatial frequency spectrum of the initial guess's error
$\h^{(0)} - \h^\ast$:  If the error has only low-spatial-frequency
components (in a sense to be clarified below), then Newton's method
has a large radius of convergence, i.e.~it will converge even for
a rather inaccurate initial guess.  However, {\em if the error has
significant high-spatial-frequency components, then we find that
Newton's method has a very small radius of convergence, i.e.~it often
fails to converge even when the error $\h^{(0)} - \h^\ast$ is very
small\/}.

This behavior is {\em not\/} an artifact of insufficient resolution
in the finite difference grid.  Rather, it appears to be caused by a
strong nonlinearity in the continuum $H(h)$ function for high-spatial-frequency
components in $h$.  In this context there's no sharp demarcation
between ``low'' and ``high'' spatial frequencies, but in practice we
use the terms to refer to angular Fourier components varying as (say)
$\cos m\theta$ with $m \ltsim 4$ and $m \gtsim 8$ respectively.
%
%%%%%%%%%%%%%%%%%%%%
%
\subsubsection{An Example}

As an example of this behavior, consider Kerr spacetime with
dimensionless angular momentum $a \equiv J/M^2 = 0.6$.  We use the
Kerr slicing, where the time coordinate is defined by requiring
$t + r$ to be an ingoing null coordinate.  (These slices generalize
the Eddington-Finkelstein slices of Schwarzschild spacetime, and are
similarly nonmaximal, with $K$ nonzero and spatially variable throughout
the slices.)  Taking the black hole to be of dimensionless unit mass,
the (only) apparent horizon in such a slice is the coordinate sphere
$r = h^\ast(\theta,\phi) = 1 + \sqrt{1 - a^2} = 1.8$.

For this example we consider two different initial guesses for
the horizon position: one containing only low-spatial-frequency
errors, $r = h^{(0)}(\theta,\phi) = 1.8 + 0.1 \cos 4 \theta$, and one
containing significant high-spatial-frequency errors,
$r = h^{(0)}(\theta,\phi) = 1.8 + 0.1 \cos 10 \theta$.  Notice that
both initial guesses are quite close to the actual horizon shape,
differing from it by slightly less than 5\%.  We use a finite
difference grid with $\Delta \theta = \frac{\pi/2}{50}$, which
is ample to resolve all the trial horizon surfaces occurring in
the example.

Figure~\hbox{\ref{fig-Kerr-hp4-hp10}(a)} shows the behavior of
Newton's method for the low-spatial-frequency-error initial guess.
As can be seen, here Newton's method converges without difficulty.

Figure~\hbox{\ref{fig-Kerr-hp4-hp10}(b)} shows the behavior of
Newton's method for the high-spatial-frequency-error initial guess.
Here Newton's method fails to converge: the successive iterates
(trial horizon surfaces) $\h^{(k)}$ move farther and farther away
the horizon, and rapidly become more and more nonspherical.

Figure~\hbox{\ref{fig-Kerr-hp4-hp10}(c)} shows the behavior of the
modified Newton's method for this same high-spatial-frequency-error
initial guess.  Although the first iteration still moves the trial
horizon surface somewhat inward from the horizon, the nonsphericity
damps rapidly, and the successive iterates (trial horizon surfaces)
quickly converge to the horizon.

Notice that all the intermediate iterates (trial horizon surfaces)
in this example are well-resolved by the finite difference grid.  To
verify that insufficient grid resolution isn't a factor in the behavior
of the horizon finder here, we have rerun all three parts of this example
with several higher grid resolutions, obtaining results essentially
identical to those plotted here.

More quantitatively, following our discussion of the Newton-Kantorovich
method in section~\ref{sect-Newton-Kantorovich-method}, we have made
3-grid convergence tests of each intermediate iterate (trial horizon
surface) in this example.  For example, figure~\ref{fig-Kerr-hp10-Newton-conv}
shows a 3-grid convergence test for the Newton iterate (trial horizon
surface) $\h^{(2)}$ plotted in figure~\hbox{\ref{fig-Kerr-hp4-hp10}(b)},
using grids with resolutions
$\Delta \theta
	= \frac{\pi/2}{50}$:$\frac{\pi/2}{100}$:$\frac{\pi/2}{200}$.
Notice that despite the iteration-divergence of the Newton iteration,
this iterate shows excellent 4th~order finite-difference-convergence.
The other Newton and modified-Newton iterates (trial horizon surfaces)
in our example all similarly show excellent 4th~order
finite-difference-convergence.

We conclude that the iteration-divergence of Newton's method seen in
figure~\hbox{\ref{fig-Kerr-hp4-hp10}(b)}, is in fact an inherent property
of the continuum Newton-Kantorovich algorithm for this initial guess
and slice.  Looking at the internal structure of this algorithm, we
see that its only approximation is the linearization of the continuum
$H(h)$ function in \eqref{eqn-continuum-H(h)-linearized}, so the
algorithm's iteration-divergence must (can only) be due to nonlinearity
in the continuum $H(h)$ function.
%
%%%%%%%%%%%%%%%%%%%%
%
\subsubsection{The Horizon-Perturbation Survey}

To investigate how general the poor convergence of Newton's method
seen in this example is, and to what extent it also occurs for the
modified Newton's method, we have made a Monte Carlo numerical survey
of both algorithms' behavior over a range of different initial-guess-error
spatial frequency spectra.

For this survey we first fix a particular horizon-finding algorithm.
Suppose we are given a slice containing an apparent horizon at the
continuum position $h^\ast$, and consider running the horizon finder
with the generic perturbed initial guess
\begin{equation}
h = h^\ast
    + \sum_{{\scriptstyle m = 0} \atop {\scriptstyle \text{$m$ even}}}
	  ^M
      c_m \cos m\theta
\end{equation}
for some set of initial-guess-error Fourier coefficients $\{ c_m \}$.
(Here we include only even-$m$ cosine terms in $\theta$ so as to
preserve axisymmetry and equatorial reflection symmetry, which our
code requires.)

For each value of $M$ we define the horizon finder's
\defn{convergence region} in $\{ c_m \}$-space to be the set of
coefficients $\{ c_m \}$ for which the horizon finder converges
(we presume to the correct solution).  For example, the convergence
region will in practice always include the origin in $\{ c_m \}$-space,
since there $h = h^\ast$, so the initial guess differs from the
exact solution of the discrete $\H(\h) = 0$ equations only by the
small $\H(\h)$ finite differencing error.

We define $V_M$ to be the (hyper)volume of the convergence region.
As described in detail in appendix~\ref{app-hps-details}, we estimate
$V_M$ by Monte Carlo sampling in $\{ c_m \}$-space.  Given $V_M$, we
then define the \defn{volume ratio}
\begin{equation}
R_M = \left\{
       \begin{array}{l@{\quad}l}
       V_0		& \text{if $M = 0$}	\\[1ex]
       {\displaystyle \frac{V_M}{V_{M-2}}}
			& \text{if $M \geq 2$}	%%%\\
       \end{array}
       \right.
					\, \text{,}	% text punctuation
									%%%\\
\end{equation}
so that $R_M$ measures the average radius of convergence of the
horizon finder in the $c_M$ dimension.
%
%%%%%%%%%%%%%%%%%%%%
%
\subsubsection{Results and Discussion}

We have carried out such a horizon-perturbation survey for the same
Kerr slices of the unit-mass spin-$0.6$ Kerr spacetime used in the
previous example, for both the Newton and the modified-Newton algorithms,
for $M = 0$, $2$, $4$, \dots, $12$.  Figure~\ref{fig-Kerr-hps} shows
the resulting volume ratios.  Although the precise values are somewhat
dependent on the details of our implementation and on the test setup
(in particular on the position of the inner grid boundary, which is
at $r = 1$ for these tests), the relative trends in the data should
be fairly generic.  These tests use a grid with
$\Delta \theta = \frac{\pi/2}{50}$, which is adequate to resolve all
the perturbed trial horizon surfaces.

As can be seen from the figure, the modified-Newton algorithm is
clearly superior to the Newton algorithm, increasing the radius of
convergence by a factor of $2$--$3$ at high spatial frequencies.
However, both algorithms' radia of convergence still fall rapidly
with increasing spatial frequency, approximately as $1 / M^{3/2}$,
although the rate is slightly slower for the modified-Newton than
for the Newton algorithm.  The radius of convergence of Newton's
method falls below~0.1 ($\sim \! 5\%$~of the horizon radius)
by~$M \gtsim 10$, and the data suggest that the radius of convergence
of the modified-Newton method would be similarly small by~$M \gtsim 18$.

Since the grid resolution is adequate, we again conclude that the
small radius of convergence of Newton's method must be due to a
strong high-spatial-frequency nonlinearity in the continuum $H(h)$
function.  Our horizon-perturbation survey covers only a single
axisymmetric initial slice and generic axisymmetric perturbations of
the initial guess, but it seems unlikely that the nonlinearity would
diminish for more general cases.  Huq (Ref.~\cite{Huq-95}) has made
limited tests with nonaxisymmetric spacetimes and high-spatial-frequency
perturbations, and has found (poor) convergence of Newton's method
similar to our results.

Although we write the continuum horizon function as $H = H(h)$, it's
more accurate to write this as $H = H(g_{ij},K_{ij},h)$, since $H$ also
depends on the slice's field variables and their spatial derivatives.
Examining the functional form of the $H(g_{ij},K_{ij},h)$ function
in~\eqrefs{eqn-3H(ABCD)} and~\eqrefb{eqn-ABCD(h)}, we see that $H$
depends on the $g^{ij}$ components in a manner broadly similar to its
dependence on $h$.  We thus conjecture that the $H(g_{ij},K_{ij},h)$
function may exhibit strong high-spatial-frequency nonlinearity in the
field variables, in particular in the $g^{ij}$ components, similar to
its nonlinear dependence on $h$.

If this is the case, then high-spatial-frequency variations in the
field variables, such as would be caused by high-frequency gravitational
radiation, might well impair the convergence of Newton's method in
a manner similar to high-spatial-frequency perturbations in $h$.
Further investigation of this possibility, either by analytical
study of the nonlinear structure of the $H(g_{ij},K_{ij},h)$ function,
or by numerical investigations, would be very interesting.  Fortunately,
however, those (few) dynamic black hole spacetimes which have been
explicitly computed thus far (for example Ref.~\cite{AHSSS-95}) seem
to contain mainly low-frequency gravitational radiation.

In general, how serious a problem is the poor high-spatial-frequency
convergence of Newton's method?  Given a sufficiently good initial
guess, Newton's method still converges very rapidly (quadratically),
so the key question is, how good is the initial guess in practice?
Two cases seem to be of particular importance:  If the horizon finder
is being used to update a horizon's position at each time step of a
time evolution, then the previous time step's horizon position probably
provides a sufficiently good initial guess for Newton's method to
converge well.  In contrast, if the horizon finder is being used on
initial data, or in a time evolution where there is no nearby horizon
in the previous time step, then significant initial-guess errors can
be expected, and Newton's method may converge poorly.
%
%%%%%%%%%%%%%%%%%%%%%%%%%%%%%%%%%%%%%%%%%%%%%%%%%%%%%%%%%%%%%%%%%%%%%%%%%%%%%%%
%
\section{Accuracy of the Horizon Finder}
\label{sect-accuracy}

We now consider the accuracy of the Newton's-method horizon finding
algorithm.  That is, assuming the Newton or modified-Newton iteration
converges, how close is the horizon finder's final numerically
computed horizon position to the (a) true continuum horizon position
$h^\ast$?

The horizon finder computes Newton or modified-Newton iterates
(trial horizon surfaces) $\h^{(k)}$ for $k = 0$, $1$, $2$, \dots,
until some convergence criterion is satisfied, say at $k = p$.
Because of the extremely rapid convergence of the Newton and
modified-Newton iterations once the error is sufficiently small
(cf.~section~\ref{sect-Newton's-method}), there's little extra cost
in using a very strict convergence criterion, i.e.~in solving the
discrete $\H(\h) = 0$ equations to very high accuracy.  In our horizon
finder we typically require $\| \H(\h^{(p)}) \|_\infty < 10^{-10}$.

We denote the exact solution of the discrete $\H(\h) = 0$ equations
by $\h^\ast$.  Given that $\| \H(\h^{(p)}) \|$ is reasonably small,
then from standard matrix-perturbation theory (see, for example,
Refs.~\cite{Linpack-book-conditioning,Golub-Van-Loan-conditioning}),
$\| \h^{(p)} - \h^\ast \| \ltsim \kappa \| \H(\h^{(p)}) \|$, where
$\kappa$ is the condition number of the (presumably nonsingular)
Jacobian matrix $\Jac[\H(\h)]$ at the horizon position.

If we take the convergence tolerance to be strict enough for
$\| \h^{(p)} - \h^\ast \|$ to be negligible, then the overall
accuracy of the horizon finder, i.e.~the external error
$\| \h^{(p)} - h^\ast \|$ in the computed horizon position, is thus
limited only by the closeness with which the discrete $\H(\h) = 0$
equations approximate the continuum $H(h) = 0$ equations, i.e.~by
the accuracy of the $\H(\h)$ finite differencing.  This potential
for very high accuracy is one of the main advantages of the
Newton's-method horizon-finding algorithm.

For an example of the accuracy attainable in practice, we again
consider the Kerr slices of the unit-mass spin-$0.6$ Kerr spacetime.
However, to make the horizon deviate from a coordinate sphere and
hence be a more significant test case for our horizon finder, we
apply the spatial coordinate transformation
\begin{mathletters}
						\label{eqn-w4Kerr-coord-xform}
\begin{equation}
r \to r + \frac{b^2}{b^2 + r^2}
	  \Bigl( a_2 \cos 2 \theta + a_4 \cos 4 \theta \Bigr)
\end{equation}
to the slice, where the parameters are given by
\begin{equation}
b = 5		\qquad	a_2 = 0.75	\qquad	a_4 = 0.05
					\, \text{.}	% text punctuation
\end{equation}
\end{mathletters}
As shown in figure~\hbox{\ref{fig-w4Kerr}(a)}, in the transformed
coordinates this gives a strongly non-spherical ``peanut-shaped''
horizon, similar in shape to those around a pair of coalescing
black holes.

We have run our horizon finder on this slice, using the
warped-coordinate coordinate sphere $r = 1.8$ as an initial guess
and a grid resolution of $\Delta \theta = \frac{\pi/2}{50}$.  We
used the modified-Newton algorithm, which converged to the horizon
without difficulty.  (The convergence took 9~iterations, but would
have taken only 6~iterations in the absence of our 10\% restriction
on the relative change in any component of $\h$ in a single outer
iteration, cf.~section~\ref{sect-modifications-of-Newton's-method}.)
Figure~\hbox{\ref{fig-w4Kerr}(a)} shows the initial guess and the
final numerically computed horizon position.

Figure~\hbox{\ref{fig-w4Kerr}(b)} shows the results of a 2-grid
convergence test of the final numerically computed horizon position
for this example, using grids with resolutions
$\Delta \theta = \frac{\pi/2}{50}$:$\frac{\pi/2}{100}$.  As can be
seen, the numerically computed solution shows excellent 4th~order
convergence.  Moreover, the numerically computed horizon positions
are very accurate, with $\| \h^{(p)} - h^\ast \| \sim 10^{-5} (10^{-6})$
for a grid resolution of
$\Delta \theta = \frac{\pi/2}{50} (\frac{\pi/2}{100})$.  Errors of
this magnitude are typical of what we find for Newton's-method horizon
finding using 4th~order finite differencing, so long as the grid
adequately resolves the horizon shape.
%
%%%%%%%%%%%%%%%%%%%%%%%%%%%%%%%%%%%%%%%%%%%%%%%%%%%%%%%%%%%%%%%%%%%%%%%%%%%%%%%
%
\section{Finding {\em Outermost\/} Apparent Horizons}
\label{sect-finding-outermost-apparent-horizons}

The main focus of this paper is on locally finding apparent horizons,
i.e.~on finding an apparent horizon in a neighborhood of the initial
guess.  However, there's a related global problem of some interest
which has heretofore attracted little attention, that of finding or
recognizing the {\em outermost\/} apparent horizon in a slice.
(By \defn{recognizing} the outermost apparent horizon we mean the
problem of determining whether or not a given apparent horizon is
in fact the outermost one in a slice.)

These global problems are of particular interest when apparent
horizons are used to set the inner boundary of a black-hole-excluding
grid in the numerical evolution of a multiple-black-hole spacetime,
as discussed by Refs.~\cite{Thornburg-talk-91,Seidel-Suen-92,%%%
Thornburg-PhD,ADMSS-95}.  In this context, we can use the appearance
of a new outermost apparent horizon surrounding the previously-outermost
apparent horizons around two black holes as a diagnostic that the
black holes have collided and coalesced into a single (distorted)
black hole.  As suggested by Ref.~\cite{Thornburg-PhD}, we can then
generate a new numerical grid and attach it to the new outermost
apparent horizon, and continue the evolution on the exterior of the
new (distored) black hole.

So far as we know, no reliable algorithms are known for finding or
recognizing outermost apparent horizons in nonspherical spacetimes.
(For spherical spacetimes, a 1-dimensional search on $H(r)$ suffices.)
If started with a very large 2-sphere as the initial guess, the
curvature flow method might well converge to the outermost horizon
in the slice, but as mentioned in section~\ref{sect-algorithms-survey},
the theoretical justification for this method's convergence is only
valid in time-symmetric ($K_{ij} = 0$) slices.

For the remaining local-horizon-finding algorithms surveyed in
section~\ref{sect-algorithms-survey}, including the Newton's-method
one, we know of no better method for locating or recognizing outermost
horizons than trying the local-horizon-finder with a number of
different initial guesses near the suspected position of an outermost
horizon.  If this method succeeds it locates a horizon, but there's
still no assurance that this horizon is the outermost one in the
slice.  Moreover, if all the local-horizon-finding trials fail, this
may mean that there's no horizon in the vicinity of the initial guesses,
or it may only mean that a horizon is present nearby but the method
failed to converge to it.  It's also not clear how many local-horizon-finding
trials should be made, nor just how their initial guesses should be
chosen.

This is clearly not a satisfactory algorithm.  Further research to
develop reliable algorithms for finding or recognizing outermost
apparent horizons in generic (nonspherical, nonmaximal) slices would
be very useful.
%
%%%%%%%%%%%%%%%%%%%%%%%%%%%%%%%%%%%%%%%%%%%%%%%%%%%%%%%%%%%%%%%%%%%%%%%%%%%%%%%
%
\section{Conclusions}

We find Newton's method to be an excellent horizon-finding algorithm:
it handles fully generic slices, it's fairly easy to implement, it's
very efficient, it's generally robust in its convergence, and it's very
accurate.  These properties are all well known, and Newton's method is
widely used for horizon finding.  In this paper we focus on two key
aspects of this algorithm: the computation of the Jacobian matrix, and
the algorithm's global convergence behavior.

Traditionally, the Newton's-method Jacobian matrix is computed by a
numerical perturbation technique.  In this paper we present a much
more efficient ``symbolic differentiation'' technique.  Conceptually,
this entails differentiating the actual finite difference equations
used to compute the discrete horizon function $\H(\h)$.  However,
provided the finite differencing scheme commutes with linearization,
the computation can instead be done by first differentiating the
continuum horizon function $H(h)$, then finite differencing.  (This
is essentially just the ``Jacobian part'' of the Newton-Kantorovich
method for solving nonlinear PDEs.)

In our axisymmetric-spacetime (2-dimensional) numerical code, this
method is about a factor of~$5$ faster than than any other Jacobian
computation method.  In fact, the Jacobian computation using this
method is only $1.5$--$2$ times more expensive than the simple
evaluation of $\H(\h)$.  We expect the symbolic differentiation
method's relative advantage over other Jacobian computation methods
to be roughly similar for other axisymmetric-spacetime (2-dimensional)
codes, and an additional factor of $\sim \! 3$--$5$ larger for
fully-general-spacetime (3-dimensional) codes.

We had previously suggested (Ref.~\cite{Thornburg-PhD}) that
symbolic-differentiation Jacobian computations would be quite difficult,
necessarily requiring substantial support from a (computer) symbolic
computation system.  Several colleagues have expressed similar opinions
to us.  However, this turns out not to be the case:  we computed all
the symbolic-differentiation Jacobian coefficients for our horizon finder
by hand in only a few few pages of algebra.  Some of the coefficients
are fairly complicated, but no more so than many other computations
in $3+1$ numerical relativity.

We find the actual programming of the symbolic differentiation
Jacobian computation to be only moderately more difficult than that
of a numerical perturbation computation.  In order to be confident
of the correctness of a symbolic differentiation Jacobian computation,
we feel that it's highly desirable to program an independent numerical
perturbation method and make an end-to-end comparison of the resulting
Jacobian matrices.  The comparison Jacobian computation may be
programmed for simplicity at the expense of efficiency, so it needn't
add much to the overall symbolic-differentiation implementation effort.

Turning now to the convergence behavior of Newton's method, we
find that so long as the error in the initial guess (its deviation
from the true horizon position) contains only low-spatial-frequency
components, a Newton's-method horizon finder has a large (good)
radius of convergence, i.e.~it converges even for rather inaccurate
initial guesses.  However, if the error in the initial guess contains
significant high-spatial-frequency components, then we find that
Newton's method has a small (poor) radius of convergence, i.e.~it
may fail to converge even when the initial guess is quite close to
the true horizon position.  In this context there's no sharp demarcation
between ``low'' and ``high'' spatial frequencies, but in practice we
use the terms to refer to angular Fourier components varying as (say)
$\cos m\theta$ with $m \ltsim 4$ and $m \gtsim 8$ respectively.

Using a Monte Carlo survey of initial-guess-error Fourier-coefficient
space, we find that the radius of convergence for Newton's method
falls rapidly with increasing spatial frequency, approximately as
$1 / m^{3/2}$.  A simple ``line-search'' modification of Newton's
method roughly doubles the horizon finder's radius of convergence,
and slightly slows the rate of decline with spatial frequency.  Using
a robust nonlinear-algebraic-equations code to solve the discrete
$\H(\h) = 0$ equations would probably give some further improvement,
but we doubt that it would change the overall trend.

Using quantitative convergence tests, we demonstrate that the poor
high-spatial-frequency convergence behavior of Newton's method is
{\em not\/} an artifact of insufficient resolution in the finite
difference grid.  Rather, it appears to be inherent in the (a) strong
nonlinearity of the continuum $H(h)$ function for high-spatial-frequency
components in $h$.  We conjecture that $H$ may be similarly nonlinear
in its high-spatial-frequency dependence on the inverse-metric components.
If so, then the presence of high-frequency gravitational radiation
might well also impair the convergence of Newton's method, and possibly
other horizon-finding methods as well.  Further investigation of this
possibility would be very interesting.

Fortunately, if the horizon finder is being used to update a
horizon's position at each time step of a time evolution, then the
previous time step's horizon position probably provides a sufficiently
good initial guess for Newton's method to converge well.

Provided it converges, the Newton's-method algorithm for horizon
finding is potentially very accurate, in practice limited only by
the accuracy of the $\H(\h)$ finite differencing scheme.  Using
4th~order finite differencing, we demonstrate that the error in
the numerically computed horizon position, i.e.~the deviation of
$\h$ from the true continuum horizon position, shows the expected
$O((\Delta \theta)^4)$ scaling with grid resolution $\Delta \theta$,
and is typically $\sim \! 10^{-5} (10^{-6})$ for a grid resolution of
$\Delta \theta = \frac{\pi/2}{50} (\frac{\pi/2}{100})$.

Finally, we have argued that considerable further research is needed
to develop algorithms for finding or recognizing the {\em outermost\/}
apparent horizon in a slice.  This is an important problem for the
numerical evolution of multiple-black-hole spacetimes with the black
holes excluded from the numerical evolution, but so far as we know
no reliable algorithms are known for it except in spherical symmetry.
%
%%%%%%%%%%%%%%%%%%%%%%%%%%%%%%%%%%%%%%%%%%%%%%%%%%%%%%%%%%%%%%%%%%%%%%%%%%%%%%%
%
\section*{Acknowledgments}

We thank M.~Huq for numerous useful conversations on horizon finding,
and for helpful comments on various drafts of this paper.  We thank
D.~Bernstein for communicating unpublished research notes on the
curvature-flow method to us.  We thank W.~G.~Unruh and the University
of British Columbia Physics Department for their hospitality and the
use of their research facilities.  We thank J.~Wolfgang for major
assistance in setting up computer facilities, and G.~Rodgers for
financial support.
%
%%%%%%%%%%%%%%%%%%%%%%%%%%%%%%%%%%%%%%%%%%%%%%%%%%%%%%%%%%%%%%%%%%%%%%%%%%%%%%%
%
\appendix
%
%%%%%%%%%%%%%%%%%%%%%%%%%%%%%%%%%%%%%%%%%%%%%%%%%%%%%%%%%%%%%%%%%%%%%%%%%%%%%%%
%
\section{Symbolic-Differentiation Jacobian Coefficients}
\label{app-SD-Jac-coeffs}

In this appendix we tabulate all the nonzero symbolic-differentiation
Jacobian coefficients for $\three\! H(h)$ and its subfunctions.  These
are used in the 3d.sd.1s and 3d.sd2.2s Jacobian-computation methods.
All the coefficients are obtained by straightforward, if somewhat tedious,
linearizations in the manner of~\eqref{eqn-P(Q)-and-derivs-linearized},
starting from the defining equations noted.

For $\three\! H(h)$, starting from~\eqrefs{eqn-3H(ABCD)}
and~\eqrefb{eqn-ABCD(h)}, the coefficients are
\begin{mathletters}
\begin{eqnarray}
\frac{\partial \, \three\! H}{\partial (\partial_x h)}
	& = &	{}
		\frac{1}{D^{3/2}}
		\, \frac{\partial A}{\partial (\partial_x h)}
		+ \frac{1}{D^{1/2}}
		  \, \frac{\partial B}{\partial (\partial_x h)}
		+ \frac{1}{D}
		  \, \frac{\partial C}{\partial (\partial_x h)}
							\nonumber	\\
	&   &	\quad
		{}
		- \left(
		  \frac{3}{2} \frac{A}{D^{5/2}}
		  + \frac{1}{2} \frac{B}{D^{3/2}}
		  + \frac{C}{D^2}
		  \right)
		  \frac{\partial D}{\partial (\partial_x h)}
									\\
\frac{\partial \, \three\! H}{\partial (\partial_{xy} h)}
	& = &	{}
		\frac{1}{D^{3/2}}
		\, \frac{\partial A}{\partial (\partial_{xy} h)}
		+ \frac{1}{D^{1/2}}
		  \, \frac{\partial B}{\partial (\partial_{xy} h)}
									\\
\noalign{\hbox{where}}
\frac{\partial A}{\partial (\partial_x h)}
	& = &	{}
		- \Bigl[
		  g^{ux} (g^{vr} - g^{vw} \partial_w h)
		  + g^{vx} (g^{ur} - g^{uw} \partial_w h)
		  \Bigr]
		  \partial_{uv} h
							\nonumber	\\
	&   &	\quad
		{}
		+ \thalf
		  g^{ix} \Bigl[
			 \partial_i g^{rr}
			 - 2 (\partial_i g^{ru}) \partial_u h
			 + (\partial_i g^{uv}) (\partial_u h) (\partial_v h)
			 \Bigr]
							\nonumber	\\
	&   &	\quad
		{}
		+ (g^{ir} - g^{iu} \partial_u h)
		  \Bigl[
		  \partial_i g^{xr} - (\partial_i g^{xv}) \partial_v h
		  \Bigr]
									\\
\frac{\partial B}{\partial (\partial_x h)}
	& = &	{}
		- \partial_i g^{ix}
		- (\partial_i \ln \sqrt{g}) g^{ix}
									\\
\frac{\partial C}{\partial (\partial_x h)}
	& = &	{}
		- 2 (K^{xr} - K^{xu} \partial_u h)
									\\
\frac{\partial D}{\partial (\partial_x h)}
	& = &	{}
		- 2 (g^{xr} - g^{xu} \partial_u h)
									\\
\frac{\partial A}{\partial (\partial_{xy} h)}
	& = &	(g^{xr} - g^{xu} \partial_u h) (g^{yr} - g^{yu} \partial_u h)
									\\
\frac{\partial B}{\partial (\partial_{xy} h)}
	& = &	- g^{xy}
					\, \text{.}	% text punctuation
									%%%\\
\end{eqnarray}
\end{mathletters}

For $s_i(h)$, starting from~\eqref{eqn-s-d(h)}, the coefficients are
\begin{eqnarray}
\frac{\partial s_u}{\partial (\partial_x h)}
	& = &	\left\{
		\begin{array}{l@{\quad}l}
		1	& \text{if $u = x$}	\\
		0	& \text{otherwise}	%%%\\
		\end{array}
		\right.
					\, \text{.}	% text punctuation
									%%%\\
\end{eqnarray}

For $n^i(h)$, starting from~\eqref{eqn-n-u(h)}, the coefficients are
\begin{eqnarray}
\frac{\partial n^i}{\partial (\partial_x h)}
	& = &	{}
		- \frac{g^{ix}}{D^{1/2}}
		+ \frac{
		       (g^{ir} - g^{iu} \partial_u h)
		       (g^{xr} - g^{xv} \partial_v h)
		       }
		       {D^{3/2}}
					\, \text{.}	% text punctuation
									%%%\\
							\label{eqn-Jac[n-u(h)]}
\end{eqnarray}

For $n^i(s_u)$, starting from~\eqref{eqn-n-u(s-d)}, the coefficients
are
\begin{eqnarray}
\frac{\partial n^i}{\partial s_u}
	& = &	\frac{g^{iu}}{(g^{kl} s_k s_l)^{1/2}}
		- \frac{(g^{ik} s_k) (g^{ul} s_l)}{(g^{kl} s_k s_l)^{3/2}}
					\, \text{.}	% text punctuation
									%%%\\
\end{eqnarray}

For $\three\! H(s_i)$, starting from~\eqrefs{eqn-3H(ABCD)}
and~\eqrefb{eqn-ABCD(s-d)}, the coefficients are
\begin{mathletters}
\begin{eqnarray}
\frac{\partial \, \three\! H}{\partial s_x}
	& = &	{}
		\frac{1}{D^{3/2}} \, \frac{\partial A}{\partial s_x}
		+ \frac{1}{D^{1/2}} \, \frac{\partial B}{\partial s_x}
		+ \frac{1}{D} \, \frac{\partial C}{\partial s_x}
							\nonumber	\\
	&   &	\quad
		{}
		- \left(
		  \frac{3}{2} \frac{A}{D^{5/2}}
		  + \frac{1}{2} \frac{B}{D^{3/2}}
		  + \frac{C}{D^2}
		  \right)
		  \frac{\partial D}{\partial s_x}
									\\
\frac{\partial \, \three\! H}{\partial (\partial_x s_y)}
	& = &	{}
		\frac{1}{D^{3/2}}
		\, \frac{\partial A}{\partial (\partial_x s_y)}
		+ \frac{1}{D^{1/2}}
		  \, \frac{\partial B}{\partial (\partial_x s_y)}
									\\
\noalign{\hbox{where}}
\frac{\partial A}{\partial s_x}
	& = &	{}
		- \Bigl[ g^{ix} (g^{jk} s_k) + g^{jx} (g^{ik} s_k) \Bigr]
		  \partial_i s_j
							\nonumber	\\
	&   &	\quad
		{}
		- \thalf g^{ix} \Bigl[ (\partial_i g^{kl}) s_k s_l \Bigr]
		- (g^{ij} s_j) \Bigl[ (\partial_i g^{xk}) s_k \Bigr]
									\\
\frac{\partial B}{\partial s_x}
	& = &	{}
		- \partial_i g^{ix}
		+ (\partial_i \ln \sqrt{g}) g^{ix}
									\\
\frac{\partial C}{\partial s_x}
	& = &	{}
		2 K^{xi} s_i
									\\
\frac{\partial D}{\partial s_x}
	& = &	{}
		2 g^{xi} s_i
									\\
\frac{\partial A}{\partial (\partial_x s_y)}
	& = &	- (g^{xk} s_k) (g^{yl} s_l)
									\\
\frac{\partial B}{\partial (\partial_x s_y)}
	& = &	g^{xy}
					\, \text{.}	% text punctuation
									%%%\\
\end{eqnarray}
\end{mathletters}

For $\three\! H(n^i)$, starting from~\eqref{eqn-3H(n-u)}, the
coefficients are
\begin{mathletters}
\begin{eqnarray}
\frac{\partial \, \three\! H}{\partial n^x}
	& = &	\partial_x \ln \sqrt{g} + 2 K_{xi} n^i
									\\
\frac{\partial \, \three\! H}{\partial (\partial_x n^y)}
	& = &	\left\{
		\begin{array}{l@{\quad}l}
		1	& \text{if $x = y$}	\\
		0	& \text{otherwise}	%%%\\
		\end{array}
		\right.
					\, \text{.}	% text punctuation
									%%%\\
\end{eqnarray}
\end{mathletters}
%
%%%%%%%%%%%%%%%%%%%%%%%%%%%%%%%%%%%%%%%%%%%%%%%%%%%%%%%%%%%%%%%%%%%%%%%%%%%%%%%
%
\section{Details of our Horizon-Finding Code}
\label{app-code-details}

In this appendix we outline those details of our horizon-finding
code relevant to the remainder of this paper.

Our horizon finder implements all the horizon-function and
Jacobian-computation methods discussed in this paper, as summarized
in table~\ref{tab-methods-comparison}.  It's part of a larger $3+1$
code under development, designed to time-evolve an asymptotically
flat axisymmetric vacuum spacetime containing a single black hole
present in the initial data.  The black hole is excluded from the
numerical grid in the manner described by Refs.~\cite{Thornburg-talk-91,%%%
Seidel-Suen-92,Thornburg-PhD,ADMSS-95}.  The code uses 4th~order
centered finite differencing (5-point molecules) for finite differencing,
on a 2-dimensional polar-spherical-coordinate grid.  (The code also
assumes equatorial reflection symmetry, but this is merely for convenience
and could easily be changed.)  The code uses a ``PDE compiler'' to
automatically generate all the finite differencing and other
grid-computation code, including that for the horizon function and
Jacobian computations, from a high-level tensor-differential-operator
specification of the $3+1$ equations.

The entire code is freely available on request from the author,
and may be modified and/or redistributed under the terms of the
GNU Public License.  The code should be easily portable to any
modern computing platform.  It's mainly written in ANSI~C (about
30K~lines) and the Maple symbolic-computation language (about
9K~lines for the PDE compiler itself, and about 6K~lines for the
$3+1$ equations), together with about 1K~lines of awk.  The code
for the horizon finder itself is about 6K~lines of C and 2K~lines
of Maple, but a large part of this is due to its supporting many
different combinations of finite differencing schemes and
horizon-function and Jacobian computation methods.  We estimate
that an implementation supporting only a single differencing scheme
and horizon-function and Jacobian computation method, supplemented
by a not-optimized-for-efficiency independent Jacobian computation
for debugging purposes (cf.~section~\ref{sect-methods-comparison}),
would be a factor of $\sim \! 4$ smaller.

The code takes the metric, extrinsic curvature, and other $3+1$
field tensors to be algebraically fully general, i.e.~it permits
all their coordinate components to be nonzero.  To avoid $z$~axis
coordinate singularities, the code uses a hybrid of polar spherical
and Cartesian coordinates as a tensor basis.  As discussed in detail
by Ref.~\cite{Thornburg-PhD}, for the subset of the slice containing
the code's (2-dimensional) grid, this hybrid coordinate system
combines the convenient topology of polar spherical coordinates with
the singularity-free nature of Cartesian coordinates.

For present purposes, the key consequence of this $z$-axis-handling
method is that in this work we've made no effort to avoid expressions
which would be singular on the $z$~axis if polar spherical coordinates
were used as a tensor basis.  We haven't investigated this case in
detail, but we suspect such singularities would be widespread.
%
%%%%%%%%%%%%%%%%%%%%%%%%%%%%%%%%%%%%%%%%%%%%%%%%%%%%%%%%%%%%%%%%%%%%%%%%%%%%%%%
%
\section{Our Sparse-Jacobian Storage Scheme}
\label{app-sparse-Jacobian-storage}

As mentioned in section~\ref{sect-methods-comparison}, all the
Jacobian matrices involved in horizon finding are highly sparse,
and for reasonable efficiency this sparsity {\em must\/} be exploited
in storing and computing the Jacobians.  In this appendix we briefly
describe our sparse-Jacobian storage scheme.  This scheme stores
the Jacobian by rows, and is applicable to all of the Jacobian
matrices which arise in our horizon-finding algorithm.

We consider first the storage of $\Jac[\two \H(\h)]$.  Which elements
in a specified row~$\I$ of this Jacobian are nonzero?  From the basic
definition~\eqref{eqn-Jac[2H(h)]}, we see that the nonzero elements~$\J$
are precisely those where $\two \H_\I$ depends on $\h_\J$, i.e.~those
for which $\h_\J$ enters into the computation of $\two \H_\I$.  That is,
for a 1(2)-stage $\three \H(\h)$ computation, the nonzero-Jacobian~$\J$
values are precisely those within 1(2)~molecule radia of $\I$.  This
makes it easy to store the Jacobian: for each grid point $\I$, we
simply store a molecule-sized (twice-molecule-sized) array of Jacobian
elements.

In practice, for an axisymmetric-spacetime (2-dimensional) code,
where $\I$ and~$\J$ are both 1-dimensional ($\theta$) grid indices
and the Jacobian is a band matrix, we would store the Jacobian as a
2-dimensional array with indices $\I$ and~$\J-\I$.  For a
fully-general-spacetime (3-dimensional) code, where $\I$ and~$\J$
are both 2-dimensional ($\theta$ and $\phi$) grid indices, we would
store the Jacobian as a 4-dimensional array with indices $\I_\theta$,
$\I_\phi$, $\J_\theta-\I_\theta$, and $\J_\phi-\I_\phi$, where we
temporarily use subscripts for coordinate components, and where for
pedagogical simplicity we ignore the artificial grid boundaries at
$\theta = \{0, \pi\}$ and $\phi = \{0, 2\pi\}$.

A similar storage scheme may be used for more complicated Jacobians.
For example, consider the storage of $\Jac[\three \H(\h)]$.  Here
$\I$~is a 3-dimensional grid point index for $\three \H$, while
$\J$~is a 2-dimensional grid point index for $\h$.  For a 1(2)-stage
$\three \H(\h)$ computation, the nonzero Jacobian elements in a
specified Jacobian row~$\I$ are now precisely those~$\J$ within
1(2)~angular molecule radia of the angular components of~$\I$.  Thus
for an axisymmetric-spacetime (2-dimensional) code we would store
this Jacobian as a 3-dimensional array with indices $\I_r$, $\I_\theta$,
and $\J_\theta-\I_\theta$, while for a fully-general-spacetime
(3-dimensional) code we would store the Jacobian as a 5-dimensional
array with indices $\I_r, \I_\theta$, $\I_\phi$, $\J_\theta-\I_\theta$,
and $\J_\phi-\I_\phi$.

Notice that with this storage scheme the Jacobian's structure,
i.e.~the location of its nonzero elements, is stored implicitly.
This makes this scheme considerably more efficient in both space and
time than generic ``sparse matrix'' storage schemes (for example
those of Refs.~\cite{George-Liu-81,ILUCG-85}), which invariably require
the storage of large integer or pointer arrays to record a sparse
matrix's structure.
%
%%%%%%%%%%%%%%%%%%%%%%%%%%%%%%%%%%%%%%%%%%%%%%%%%%%%%%%%%%%%%%%%%%%%%%%%%%%%%%%
%
\section{Details of our Horizon-Perturbation Survey}
\label{app-hps-details}

In this appendix we describe our Monte Carlo horizon-perturbation
survey (cf.~section~\ref{sect-global-conv-HSF-errors}) in more detail.
Given the maximum initial-guess-error spatial frequency $M$, the goal
of the survey procedure is to estimate $V_M$, the (hyper)volume in
$\{ c_m \}$-space of the horizon finder's convergence region.

To do this, we first start from the origin in $\{ c_m \}$-space, and
search outwards along each $c_m$ axis until we find coefficients for
which the horizon finder fails to converge.  This gives the intersection
of the $c_m$ coordinate axes with the boundary of the convergence
region.

We then construct a sequence of nested hypercubes (strictly speaking,
hyper-par\-a\-llel\-e\-pipeds) $C_1$, $C_2$, $C_3$, \dots{} in
$\{ c_m \}$-space, starting with $C_1$ just containing the
$c_m$-co\-or\-di\-nate-axis boundaries of the convergence region,
and expanding outwards.  We use the obvious Monte Carlo sampling
algorithm to estimate the volume of the convergence region contained
within the first hypercube $C_1$, and then within the differences
$C_{k+1} - C_k$ of the succeeding hypercubes.  We continue this
process until one of the differences contains no convergence-region
volume.  We include one additional hypercube in the sequence after
this, typically \hbox{$25$--$50$\%} larger than the previous one
in each dimension, to provide a safety margin against missing
disconnected ``islands'' or fractal zones near the boundary of the
convergence region.  (These are quite plausible; recall that the
(fractal) Julia set is just the convergence region of a simple
Newton's-method iteration.)  Finally, we compute an estimate for
$V_M$ by simply adding the convergence-region-volume estimates for
$C_1$ and each $C_{k+1} - C_k$.

Unfortunately, as $M$ and hence the dimensionality of $\{ c_m \}$-space
increases, we find that the fraction of the hypercubes and hypercube
differences occupied by the convergence region decreases rapidly, so
a very large number of horizon-finding trials is needed to obtain a
reasonable statistical accuracy for $V_M$.  (For example, the $M = 12$
points in figure~\ref{fig-Kerr-hps} required 15,000 trials each.)
It's this effect which ultimately limits the maximum value of $M$
attainable in practice by a survey of this type.
%
%%%%%%%%%%%%%%%%%%%%%%%%%%%%%%%%%%%%%%%%%%%%%%%%%%%%%%%%%%%%%%%%%%%%%%%%%%%%%%%
%

%
%%%%%%%%%%%%%%%%%%%%%%%%%%%%%%%%%%%%%%%%%%%%%%%%%%%%%%%%%%%%%%%%%%%%%%%%%%%%%%%
%
% Figure caption showing various $H(h)$ computation methods
%
\begin{figure}
%
%%%%%%%%%%%%%%%%%%%%%%%%%%%%%%%%%%%%%%%%
%
\caption[$\H(\h)$ Computation Methods]
	{%%%
	This figure illustrates the various 2-stage and 1-stage
	computation methods for the horizon function $\H(\h)$.  The
	solid arrows denote finite differencing operations, the
	dotted arrow denotes an algebraic computation, and the
	dashed arrow denotes a radial interpolation to the horizon
	position $r = \h(\theta,\phi)$.  Each path from $\h$ to $\H$
	represents a separate computation method.  Notice that there
	are three distinct 2-stage methods (using the upper arrows
	from $\two \h$ to $\three \H$ in the figure) and one 1-stage
	method (using the lower arrow from $\two \h$ to $\three \H$).%%%
	}%%%
\label{fig-H(h)-methods}
%
%%%%%%%%%%%%%%%%%%%%%%%%%%%%%%%%%%%%%%%%
%
\end{figure}
%
%%%%%%%%%%%%%%%%%%%%%%%%%%%%%%%%%%%%%%%%%%%%%%%%%%%%%%%%%%%%%%%%%%%%%%%%%%%%%%%
%
% Figure caption showing commutativity of linearization and finite differencing
%
\begin{figure}
%
%%%%%%%%%%%%%%%%%%%%%%%%%%%%%%%%%%%%%%%%
%
\caption[Commutativity of Linearization and Finite Differencing]
	{%%%
	This commutative diagram illustrates the two different
	ways a Jacobian matrix can be computed.  Given a nonlinear
	continuum function $P(Q)$, the Jacobian matrix $\Jac[\P(\Q)]$
	is logically defined in terms of the lower-left path in
	the diagram, i.e.~it's defined as the Jacobian of a nonlinear
	discrete (finite difference) approximation $\P(\Q)$ to $P(Q)$.
	However, if the operations of discretization (finite differencing)
	and linearization commute, we can instead compute the Jacobian
	by the upper-right path in the diagram, i.e.~by first
	linearizing the continuum $P(Q)$ function, then discretizing
	(finite differencing) this linearization $\delta P(\delta Q)$.
	}%%%
\label{fig-linearize-vs-FD}
%
%%%%%%%%%%%%%%%%%%%%%%%%%%%%%%%%%%%%%%%%
%
\end{figure}
%
%%%%%%%%%%%%%%%%%%%%%%%%%%%%%%%%%%%%%%%%%%%%%%%%%%%%%%%%%%%%%%%%%%%%%%%%%%%%%%%
%
% Figure caption showing Schwarzschild-spacetime H(r)
%
\begin{figure}
%
%%%%%%%%%%%%%%%%%%%%%%%%%%%%%%%%%%%%%%%%
%
\caption[$H(r)$ for Schwarzschild Spacetime]
	{%%%
	This figure shows $H(r)$ for spherical trial horizon surfaces
	with coordinate radius $r$ in an Eddington-Finkelstein slice
	of a unit-mass Schwarzschild spacetime.  Notice that for
	$r > r^{\max} \approx 4.372$, $H > 0$ and $dH / dr < 0$,
	so Newton's method diverges in this region.
	}%%%
\label{fig-Schw-H}
%
%%%%%%%%%%%%%%%%%%%%%%%%%%%%%%%%%%%%%%%%
%
\end{figure}
%
%%%%%%%%%%%%%%%%%%%%%%%%%%%%%%%%%%%%%%%%%%%%%%%%%%%%%%%%%%%%%%%%%%%%%%%%%%%%%%%
%
% figure caption showing convergence of basic vs modified Newton's method
% when the initial guess contains low- vs high-spatial-frequency errors
%
\begin{figure}
%
%%%%%%%%%%%%%%%%%%%%%%%%%%%%%%%%%%%%%%%%
%
\caption[Low- versus High-Spatial-Frequency Convergence of the Horizon Finder]
	{%%%
	This figure illustrates how the convergence behavior of
	the basic and modified Newton iterations depends on the
	spatial-frequency spectrum of the initial guess's error
	$\h^{(0)} - \h^\ast$.  In each part of the figure, the
	true continuum horizon $h^\ast$ is plotted as a solid line,
	while the horizon finder's first few iterates (trial horizon
	surfaces) $\h^{(k)}$ are plotted with dots at the grid points.
	Part~(a) of the figure shows the behavior of Newton's method
	for an initial-guess-error containing only low spatial
	frequencies, part~(b) shows the behavior of Newton's method
	for an initial-guess-error containing significant high
	spatial frequencies, and part~(c) shows the behavior of
	the modified Newton iteration for the same initial guess
	as part~(b).  In parts~(a) and~(c), where the iteration
	is converging, the final iterates shown are indistinguishable
	from the true continuum horizon at the scale of the figure.
	In part~(b), where the iteration is diverging, the computed
	values for the next iterate $\h^{(3)}$ (not shown) are almost
	all far outside the scale of the figure; many of them are
	in fact negative!
	}%%%
\label{fig-Kerr-hp4-hp10}
%
%%%%%%%%%%%%%%%%%%%%%%%%%%%%%%%%%%%%%%%%
%
\end{figure}
%
%%%%%%%%%%%%%%%%%%%%%%%%%%%%%%%%%%%%%%%%%%%%%%%%%%%%%%%%%%%%%%%%%%%%%%%%%%%%%%%
%
% figure caption showing 3-grid convergence tests for
% hp10 high-spatial-frequency-error-initial-guess Newton iteration,
% 2nd Newton iterate, 50k:100k:200k resolutions
%
\begin{figure}
%
%%%%%%%%%%%%%%%%%%%%%%%%%%%%%%%%%%%%%%%%
%
\caption[Convergence Tests for Part~(b) of
	 Figure~\protect\ref{fig-Kerr-hp4-hp10}]
	{%%%
	This figure shows the results of a 3-grid covergence
	test for the 2nd-iteration Newton iterate (trial
	horizon surface) $\h^{(2)}$ plotted in
	figure~\hbox{\ref{fig-Kerr-hp4-hp10}(b)}.  The line
	has slope $\tfrac{1}{16}$, appropriate for 4th~order
	convergence.  (Recall that this line {\em isn't\/}
	fitted to the data, but is rather an a~priori prediction
	with {\em no\/} adjustable parameters.)  (The absolute
	magnitude of the errors shown here is much larger
	than is typical for our horizon finder, due to a
	combination of the compounding of smaller errors
	in the earlier Newton iterate $\h^{(1)}$, and the
	very strong angular variation in both iterates
	$\h^{(1)}$ and $\h^{(2)}$.)
	}%%%
\label{fig-Kerr-hp10-Newton-conv}
%
%%%%%%%%%%%%%%%%%%%%%%%%%%%%%%%%%%%%%%%%
%
\end{figure}
%
%%%%%%%%%%%%%%%%%%%%%%%%%%%%%%%%%%%%%%%%%%%%%%%%%%%%%%%%%%%%%%%%%%%%%%%%%%%%%%%
%
% Figure body showing volume ratios for Kerr horizon-perturbation survey
%
\begin{figure}
%
%%%%%%%%%%%%%%%%%%%%%%%%%%%%%%%%%%%%%%%%
%
\caption[Volume Ratios for the Kerr-spacetime Horizon Perturbation Survey]
	{%%%
	This figure shows the volume ratios $R_M$ for the
	horizon-perturbation survey.  These measure the average
	radius of convergence of the horizon finder as a function
	of the initial-guess-error's maximum spatial frequency
	$M$.  The points and solid lines show the results for the
	modified-Newton (upper) and Newton (lower) algorithms,
	with $\pm 1 \sigma$ statistical error bars from the
	Monte Carlo estimation procedure.  The dashed line shows
	an $R_M \sim 1 / M^{3/2}$ falloff.
	}%%%
\label{fig-Kerr-hps}
%
%%%%%%%%%%%%%%%%%%%%%%%%%%%%%%%%%%%%%%%%
%
\end{figure}
%
%%%%%%%%%%%%%%%%%%%%%%%%%%%%%%%%%%%%%%%%%%%%%%%%%%%%%%%%%%%%%%%%%%%%%%%%%%%%%%%
%
% Figure caption showing example of horizon-finder behavior and errors
% for Kerr field variables + m=2,4 warping (==> peanut shape)
%
\begin{figure}
%
%%%%%%%%%%%%%%%%%%%%%%%%%%%%%%%%%%%%%%%%
%
\caption[Example of the Accuracy of our Horizon Finder]
	{%%%
	This figure illustrates the accuracy of our horizon
	finder for a test case where the horizon's coordinate
	shape is strongly non-spherical.  The figure is
	plotted using the transformed radial coordinate defined
	by~\eqref{eqn-w4Kerr-coord-xform}.  Part~(a) of the
	figure shows the ``peanut-shaped'' true continuum
	horizon position $h^\ast$, plotted as a solid line,
	and the initial guess $\h^{(0)}$ and the final numerically
	computed horizon position) $\h^{(p)}$, plotted with
	dots at the grid points.  At this scale the numerically
	computed horizon position $\h^{(p)}$ is indistinguishable
	from the true continuum position $h^\ast$.  Part~(b)
	of the figure shows the results of a 2-grid convergence
	test for the numerically computed horizon position
	$\h^{(p)}$.  The line has slope $\tfrac{1}{16}$,
	appropriate for 4th~order convergence.  (Recall again
	that this line {\em isn't\/} fitted to the data, but
	is rather an a~priori prediction with {\em no\/}
	adjustable parameters.)
	}%%%
\label{fig-w4Kerr}
%
%%%%%%%%%%%%%%%%%%%%%%%%%%%%%%%%%%%%%%%%
%
\end{figure}
%
%%%%%%%%%%%%%%%%%%%%%%%%%%%%%%%%%%%%%%%%%%%%%%%%%%%%%%%%%%%%%%%%%%%%%%%%%%%%%%%
%
% Table summarizing various horizon-function and Jacobian computation methods
%
\begin{table}
%
%%%%%%%%%%%%%%%%%%%%%%%%%%%%%%%%%%%%%%%%
%
% ... next 3 lines are for "turned table" in postscript output
% \vspace*{75mm}
% \rotate{%%%
% \vbox{%%%
%
\caption[Summary of Horizon-Function and Jacobian Computation Methods]
	{%%%
	This table summarizes the various methods for computing
	the horizon function $\two \H(\h)$ and its Jacobian
	$\Jac[\two \H(\h)]$.  The ``codes'' are shorthand labels
	for referring to the various methods.  The relative CPU
	times are as measured for our implementation (described
	in appendix~\ref{app-code-details}), and are per angular
	grid point, normalized relative to the 1-stage $\two \H(\h)$
	computation.  The notation \defn{$\s_i|\n^i$} means
	whichever of $\s_i$ and/or $\n^i$ is appropriate, depending
	on the precise 2-stage method used to compute the horizon
	function.
	}%%%
\label{tab-methods-comparison}
%
%%%%%%%%%%%%%%%%%%%%%%%%%%%%%%%%%%%%%%%%
%
\begin{center}
\def\arraystretch{1.5}			% leave extra room for $\Jac[...]$
%					% table entries
\tabcolsep=1em				% prefent "overfull \hbox" from
%					% squishing horizontal space too much
%
\begin{tabular}{lllllcl}
		& Jacobian
	&							& Horizon
	&
	& Relative	& Estimated					\\
		& Computation
	&							& Function
	&
	& CPU		& Implementation				\\
Code		& Dimensions
	& Jacobian Computation Method				& Method
	& Jacobian Matrices Used
	& Time		& Effort					\\
\hline % --------------------------------------------------------------
$\H$.1s		&
	&							& 1-stage
	&
	& ${} \equiv 1$	& Moderate					\\
$\H$.2s		&
	&							& 2-stage
	&
	& 0.7		& Low						\\
\hline % --------------------------------------------------------------
2d.np.1s	& 2-dimensional
	& Numerical perturbation of $\two \H(\h)$		& 1-stage
	& $\Jac[\two \H(\h)]$
	& 6		& Low						\\
2d.np.2s	& 2-dimensional
	& Numerical perturbation of $\two \H(\h)$		& 2-stage
	& $\Jac[\two \H(\h)]$
	& 8		& Low						\\
3d.np.1s	& 3-dimensional
	& Numerical perturbation of $\three \H(\h)$		& 1-stage
	& $\Jac[\two \H(\h)]$, $\Jac[\three \H(\h)]$
	& 7		& Low -- Moderate				\\
3d.sd.1s	& 3-dimensional
	& Symbolic differentiation of $\three \H(\h)$		& 1-stage
	& $\Jac[\two \H(\h)]$, $\Jac[\three \H(\h)]$
	& 1.5		& Moderate					\\
3d.np.2s	& 3-dimensional
	& Numerical perturbation of $\three \H(\h)$		& 2-stage
	& $\Jac[\two \H(\h)]$, $\Jac[\three \H(\h)]$
	& 8		& Low -- Moderate				\\
3d.np2.2s	& 3-dimensional
	& Numerical perturbation				& 2-stage
	& $\Jac[\two \H(\h)]$, $\Jac[\three \H(\h)]$,
	& 14		& Moderate -- High				\\
		&
	& of $\s_i|\n^i(\h)$ and $\three \H(\s_i|\n^i)$		&
	& $\Jac[\s_i|\n^i(\h)]$, $\Jac[\three \H(\s_i|\n^i)]$ 
	&		&						\\
3d.sd2.2s	& 3-dimensional
	& Symbolic differentiation				& 2-stage
	& $\Jac[\two \H(\h)]$, $\Jac[\three \H(\h)]$,
	& 5		& Moderate -- High				\\
		&
	& of $\s_i|\n^i(\h)$ and $\three \H(\s_i|\n^i)$		&
	& $\Jac[\s_i|\n^i(\h)]$, $\Jac[\three \H(\s_i|\n^i)]$ 
	&		&						%%%\\
\end{tabular}
\end{center}
%
% ... next 2 lines are for "turned table" in postscript output
% }%%%
% }%%%
%
%%%%%%%%%%%%%%%%%%%%%%%%%%%%%%%%%%%%%%%%
%
\end{table}
%
%%%%%%%%%%%%%%%%%%%%%%%%%%%%%%%%%%%%%%%%%%%%%%%%%%%%%%%%%%%%%%%%%%%%%%%%%%%%%%%
%
\end{document}